\documentclass[a4paper,11pt]{article}
\pdfoutput=1

\usepackage{jheppub} 
\usepackage{graphicx,color}
\usepackage{enumitem}
\usepackage{braket}
\usepackage{slashed}
\usepackage[utf8]{inputenc}
\usepackage{hyperref}
\usepackage{float}
\usepackage{caption}
\usepackage{subfigure}
\usepackage{makecell}
\usepackage{epsfig}
\usepackage{amsmath,amssymb}
\usepackage{hyperref}
\hypersetup{
    colorlinks=flase,
    linkcolor=black,
    filecolor=magenta,      
    urlcolor=blue,
}

\hypersetup{
	colorlinks,
	citecolor=black,
	filecolor=black,
	linkcolor=blue,
	urlcolor=black
}

\newcommand{\bea}{\begin{eqnarray}}
\newcommand{\eea}{\end{eqnarray}}
\newcommand{\be}{\begin{equation}}
\newcommand{\ee}{\end{equation}}
\newcommand{\ba}{\begin{align}}
\newcommand{\ea}{\end{align}}
\newcommand{\ben}{\begin{enumerate}}
\newcommand{\een}{\end{enumerate}}
\newcommand{\bi}{\begin{itemize}}
\newcommand{\ei}{\end{itemize}}
\newcommand{\mc}{\mathcal}
\newcommand{\comments}[1]{}
\def\nn{\nonumber}

\def\bel#1{\begin{equation} \label{#1}}

\def\vo{\mathcal{V}}
\def\vov{\langle\mathcal{V}\rangle}

\def\KK{{\scriptscriptstyle KK}}
\def\SM{{\scriptscriptstyle SM}}
\def\ED3{{\scriptscriptstyle ED3}}
\def\GUT{{\scriptscriptstyle GUT}}
\def\sixD{{\scriptscriptstyle 6D}}

\def\LVS{{\scriptscriptstyle LVS}}

\def\QCD{{\scriptscriptstyle QCD}}

\def\T{{\scriptscriptstyle T}}

\def\ALP{{\scriptscriptstyle ALP}}

\title{Moduli Stabilisation and the Statistics of Axion Physics in the Landscape}

\author[a]{Igor Broeckel,}
\author[a]{Michele Cicoli,}
\author[b]{Anshuman Maharana,}
\author[b]{Kajal Singh}
\author[c]{and Kuver Sinha}
\affiliation[a]{Dipartimento di Fisica e Astronomia, Universitá di Bologna, via Irnerio 46, 40126 Bologna, Italy and INFN, Sezione di Bologna, viale Berti Pichat 6/2, 40127 Bologna, Italy}
\affiliation[b]{Harish-Chandra Research Institute, HBNI, Jhunsi, Allahabad, UP 211019, India}
\affiliation[c]{Department of Physics and Astronomy, University of Oklahoma, Norman, OK 73019, USA}

\emailAdd{igor.broeckel@bo.infn.it}
\emailAdd{michele.cicoli@unibo.it}
\emailAdd{anshumanmaharana@hri.res.in}
\emailAdd{kajalsingh@hri.res.in}
\emailAdd{kuver.sinha@ou.edu}

\abstract{String theory realisations of the QCD axion are often said to belong to the anthropic window where the decay constant is around the GUT scale and the initial misalignment angle has to be tuned close to zero.
In this paper we revisit this statement by studying the statistics of axion physics in the string landscape. We take moduli stabilisation properly into account since the stabilisation of the saxions is crucial to determine the physical properties of the corresponding axionic partners. We focus on the model-independent case of closed string axions in type IIB flux compactifications and find that their decay constants and mass spectrum feature a logarithmic, instead of a power-law, distribution. In the regime where the effective field theory is under control, most of these closed string axions are ultra-light axion-like particles, while axions associated to blow-up modes can naturally play the role of the QCD axion. Hence, the number of type IIB flux vacua with a closed string QCD axion with an intermediate scale decay constant and a natural value of the misalignment angle is only logarithmically suppressed. In a recent paper we found that this correlates also with a logarithmic distribution of the supersymmetry breaking scale, providing the intriguing indication that most, if not all, of the phenomenologically interesting quantities in the string landscape might feature a logarithmic distribution.}

\begin{document} 
\maketitle
\flushbottom

\section{Introduction}
\label{Intro}

The Peccei-Quinn mechanism is without any doubt the most elegant solution to the strong CP problem. It postulates the existence of an anomalous global $U(1)_{\rm PQ}$ symmetry which is spontaneously broken at $f_a$. The corresponding Goldstone boson is the so-called QCD axion $a$ which enjoys a continuous shift symmetry. QCD instantons lift the axionic direction and provide a minimum where CP is conserved. The QCD axion develops a mass of order $m_a \sim \Lambda_\QCD^2/f_a$ and naturally contributes to the dark matter (DM) abundance. The phenomenologically allowed window for the axion decay constant $f_a$ is given by $10^9\,{\rm GeV}\lesssim f_a\lesssim 10^{12}\,{\rm GeV}$, where the lower bound is due to astrophysical and direct observations while the upper bound comes from the requirement to avoid DM overproduction if the initial misalignment angle takes natural $\mc{O}(1)$ values. 

This scenario for the solution of the strong CP problem relies on some assumptions which have to be checked in a UV complete embedding. Some crucial questions which need to be answered are: (i) What is the origin of the axion shift symmetry?; (ii) What dynamics breaks $U(1)_{\rm PQ}$ spontaneously and sets the value of $f_a$?; (iii) Is $f_a$ related to other important physical quantities like the Planck scale $M_p$, the string scale $M_s$, the GUT scale $M_\GUT$, the Kaluza-Klein scale $M_\KK$ or the scale of soft supersymmetry breaking terms $M_{\rm soft}$?; (iv) what dynamics breaks $U(1)_{\rm PQ}$ explicitly and sets the value of $m_a$?; (v) Is $m_a$ generated by QCD instantons or by other effects?: (vi) How many axion-like particles (ALPs) can arise from UV physics?; (vii) What is the allowed range of $f_a$ and $m_a$ for these ALPs?

Several studies performed during the last 15 years revealed that string theory can provide a successful answer to many, if not all, of the previous questions \cite{Svrcek:2006yi, Conlon:2006tq, Arvanitaki:2009fg, Cicoli:2012sz}. However string theory admits a plethora of 4D solutions which goes under the name of string \emph{landscape}. Even if all 4D string vacua share some generic features about axion physics, the number of axions and the corresponding values of $f_a$ and $m_a$ take different values in different string vacua. In order to make contact with observations, it is therefore crucial to perform a statistical analysis of the distribution in the string landscape of phenomenologically relevant quantities like $f_a$ and $m_a$ which determine the axion DM abundance. 

As stressed in our recent paper \cite{Broeckel:2020fdz} where we derived the distribution of the supersymmetry breaking scale in the string landscape, these statistical studies need to be based on a solid understanding of moduli stabilisation. In the case of axion physics, the motivation is the following. We shall focus on the type IIB flux compactifications which provide a well-defined subset of the string landscape. A model-independent origin of 4D axions is provided by the higher-dimensional gauge form $C_4$ which gives rise to pseudoscalars with a continuous shift symmetry when reduced on internal 4-cycles $\Sigma_4^i$: $\theta_i = \int_{\Sigma_4^{i}} C_4$. These axions are the imaginary parts of the K\"ahler moduli $T_i =\tau_i + {\rm i}\, \theta_i$ whose real parts $\tau_i$ control the volume of $\Sigma_4^i$ in string units. Due to a combination of supersymmetry, scale invariance and the axionic shift symmetry, at tree-level each $T_i$ is a flat direction \cite{Burgess:2020qsc}. The axionic directions $\theta_i$ are lifted by instantons which preserve only a discrete shift symmetry. On the other hand, the saxions $\tau_i$ can be stabilised either at perturbative or at non-perturbative level. Let us comment on the implications of these two situations for axion physics:
\bi
\item If a given saxion $\tau$ is fixed by non-perturbative physics as in KKLT models \cite{Kachru:2003aw}, the stabilisation is at leading order supersymmetric, implying $m_\theta\sim m_\tau \sim m_{3/2}$. Given that the absence of any cosmological moduli problem requires $m_\tau\gtrsim \mc{O}(50)$ TeV \cite{deCarlos:1993wie} and $m_{3/2}$ sets the mass of the superpartners which cannot be lower than the TeV-scale, in this case the axion $\theta$ is generically very heavy, and so cannot play the role of the QCD axion \cite{Cicoli:2012sz}.

\item If $\tau$ is stabilised by perturbative physics (such as $\alpha'$ and/or string loop corrections to the K\"ahler potential), at this level of approximation $m_\tau \sim m_{3/2}$ while $m_\theta = 0$. In the regime where the effective field theory (EFT) is under control, i.e. where non-perturbative contributions are exponentially suppressed with respect to perturbative terms, instanton effects will lift $\theta$ while inducing negligible corrections to the stabilisation of $\tau$. This would produce the mass hierarchy $m_\theta \ll m_\tau \sim m_{3/2}$ which identifies $\theta$ as a promising QCD axion candidate with $\theta \simeq a/f_a$.
\ei

Besides focusing on models where $\tau$ is fixed at perturbative level, the other conditions to be checked to get a viable QCD axion are that $\theta$ couples to the QCD sector coming from stacks of D7-branes,\footnote{Notice that when the QCD sector lives on D3-branes at singularities, $C_4$-axions are eaten up by anomalous $U(1)$ and the QCD axion arises from open string modes \cite{Allahverdi:2014ppa}.} and that stringy instantons generate a mass $m_{\theta,{\rm str}}$ for $\theta$ which is smaller than the one developed by QCD instantons, i.e. $m_{\theta,{\rm str}} \ll \Lambda_\QCD^2/f_a$. 

If all these conditions are satisfied, one has still to derive the value of $f_a$ which determines all the main phenomenological properties of the QCD axion: its mass, its couplings and its contribution to the DM abundance. Depending on the topology of the 4-cycle $\Sigma_4$, $f_a$ can be either of order $M_\KK$ for bulk cycles, or of order $M_s$ for blow-up modes \cite{Conlon:2006tq, Cicoli:2012sz}. In a given moduli stabilisation framework, these two fundamental scales can be explicitly written down in terms of the underlying parameters (like the string coupling $g_s$ and the vacuum expectation value of the tree-level superpotential $W_0$) which depend on flux quanta. By exploiting the known distributions of $g_s$ and $W_0$ in the flux landscape \cite{Denef:2004ze}, one can therefore derive the distribution of $f_a$. We shall perform this analysis by focusing on the Large Volume Scenario (LVS) \cite{Balasubramanian:2005zx, Cicoli:2008va} which fixes some K\"ahler moduli at perturbative level, and find that $f_a$ features a logarithmic distribution. 

We shall consider two possible realisations of the QCD axion: ($i$) axions associated to blow-up modes, and ($ii$) axions associated to bulk cycles. In case ($i$) $f_a$ is independent on the Standard Model (SM) gauge coupling, while in case ($ii$) the decay constant is fixed around the GUT scale by the requirement of reproducing the observed visible sector gauge coupling $\alpha_\SM$. Hence, once we focus on phenomenologically relevant vacua with $\alpha_\SM^{-1}\sim \mc{O}(10$-$100)$, only axions associated to blow-up modes feature a logarithmic distribution of $f_a$. However we consider this to be the generic situation for vacua where the EFT is under control since axions from bulk cycles require an anisotropic shape of the extra dimensions which corresponds to a tuned situation for moduli stabilisation. The reason is the interplay between two conflicting conditions: the low-energy 4D EFT can be trusted only for large values of the internal Calabi-Yau (CY) manifold, while $\alpha_\SM^{-1}\sim \mc{O}(10$-$100)$ implies that the 4-cycle supporting the SM brane system cannot be too large. 

This result confirms the naive expectation that a generic 4D string model is characterised by a QCD axion with a GUT scale decay constant which would overproduce DM if the initial misalignment angle $\theta_{\rm in}$ is not tuned close to zero. However it also shows that string vacua with a QCD axion with an intermediate scale $f_a$ and an $\mc{O}(1)$ value of $\theta_{\rm in}$ are not so rare since the number of flux vacua grows with $f_a$ only as a logarithm, instead of a power-law. 

Interestingly, in \cite{Broeckel:2020fdz} we found that also the distribution of the gravitino mass in the flux landscape is logarithmic,\footnote{To be more precise, in \cite{Broeckel:2020fdz} we concluded that the distribution of the gravitino mass is logarithmic in LVS models and power-law in KKLT scenarios. However recent explicit constructions of KKLT models \cite{Demirtas:2019sip, Demirtas:2020ffz, Blumenhagen:2020ire} might indicate a logarithmic distribution also for the KKLT case.} providing the intriguing indication that most, if not all, of the phenomenologically interesting quantities in the string landscape might feature a logarithmic distribution. Once the distributions of more than one phenomenological quantity are known, it is important to look at potential correlations among them. In our case, we find that vacua with an intermediate scale $f_a$ are also characterised by TeV-scale soft-terms, as typical of LVS models \cite{Conlon:2005ki}.

Let us finally mention that a generic CY gives rise to many K\"ahler moduli in the 4D EFT. If several of them are stabilised by perturbative effects, only one of them will play the role of the QCD axion while all the others would behave as ALPs which tend to be ultra-light in the regime where the computational control over the EFT is solid. These ALPs have interesting applications to DM \cite{Hui:2016ltb}, dark radiation \cite{Cicoli:2012aq, Higaki:2012ar,Hebecker:2014gka, Cicoli:2015bpq} and astrophysics \cite{Cicoli:2014bfa, Cicoli:2017zbx}. We shall therefore derive also the distribution in the flux landscape of the decay constants, the mass spectrum and the DM contribution of stringy ALPs, finding again a logarithmic dependence. 

One may wonder whether our findings provide a trustable representation of the generic situation for axion physics in the flux landscape since they are based on the LVS framework while other moduli stabilisation mechanisms at perturbative level have been proposed \cite{Berg:2005yu}. However recent studies of the K\"ahler cone of CY manifolds with a large number of K\"ahler moduli $h^{1,1}$ revealed that, in the regime where the volume of each holomorphic curve is larger than the string scale so that the $\alpha'$ expansion is under control, the overall volume in string units grows as $\vo \gtrsim (h^{1,1})^7$ \cite{Demirtas:2018akl}. This clearly implies that for a generic CY with $h^{1,1}\sim \mc{O}(100)$, the EFT can be under control only if the moduli are fixed at $\vo\gtrsim\mc{O}(10^{14})$. Given that only LVS models yield an exponentially large CY volume which can naturally account for such a large value of $\vo$, we believe that the genericity of our results is rather robust. Ref. \cite{Cicoli:2016chb} presented an explicit LVS moduli stabilisation procedure which can lead to an exponentially large CY volume for arbitrarily large $h^{1,1}$ exploiting instantons on del Pezzo divisors and $\mc{O}(\alpha'^3)$ corrections at $\mc{O}(F^2)$ and $\mc{O}(F^4)$ (where $F$ denotes an F-term). This moduli stabilisation scenario leads to several ultra-light axions in agreement with the expectation of \cite{Demirtas:2018akl}. 

Moreover, ref. \cite{Mehta:2020kwu,Mehta:2021pwf} derived the distributions of the axion decay constants and masses for different values of $h^{1,1}$ but at a given point in the moduli space, focusing in particular on the tip of the so-called stretched K\"ahler cone, i.e. the point closest to the origin which allows to keep the EFT under control. Interestingly, they found that the mean value of $f_a$ decreases as $h^{1,1}$ increases. Our results are complementary to the ones of \cite{Mehta:2020kwu,Mehta:2021pwf} since we included moduli stabilisation and worked out the distribution of $f_a$ and $m_a$ as a function of flux quanta, i.e. moving in the moduli space at fixed $h^{1,1}$. The results of \cite{Mehta:2020kwu, Mehta:2021pwf} can be integrated with ours since they provide the boundaries of the region in moduli space where the EFT is under control and our logarithmic distributions can be trusted, i.e. our logarithmic distributions are valid for $f_a\lesssim f_{a,{\rm max}}(h^{1,1})$ or $m_a \lesssim m_{a,{\rm max}}(h^{1,1})$ with $f_{a,{\rm max}}(h^{1,1})$ and $m_{a,{\rm max}}(h^{1,1})$ as given in \cite{Mehta:2020kwu, Mehta:2021pwf} as a function of the number of K\"ahler moduli $h^{1,1}$. 

Similar considerations apply to the comparison of our findings with the ones of \cite{Halverson:2019cmy} which noticed that, in the presence of $N\gg 1$ ALPs which are effectively massless, there is just a linear combination of them which couples to photons. Ref. \cite{Halverson:2019cmy} derived the distribution of the corresponding ALP-photon coupling $g_{a\gamma\gamma}$ as a function of $N$ (with $N\sim h^{1,1}$) at a fixed point in moduli space, choosing again the tip of the stretched K\"ahler cone. For type IIB flux vacua, they found $g_{a\gamma\gamma}(N)\sim 10^{-21}\,N^4\,{\rm GeV}^{-1}$ which, according to our previous considerations, can be considered as a lower bound for a logarithmic distribution of $g_{a\gamma\gamma}$ as a function of different flux vacua at fixed $N$, when moduli stabilisation is taken into account along the lines of our paper. 

This paper is organised as follows. In Sec. \ref{sec:LVS} we discuss in depth the interplay between axion physics and moduli stabilisation. We first describe in detail an example with $h^{1,1}=4$ where the QCD axion can arise from either a bulk or a blow-up cycle, and then we discuss a more general example with arbitrarily large $h^{1,1}$. In Sec. \ref{sec:Pheno} we derive the distribution in the type IIB flux landscape of several quantities of axion physics relevant for phenomenology: decay constants, masses, DM abundance, axion couplings to gauge bosons and axion dark radiation in Fibre Inflation models \cite{Cicoli:2008gp, Cicoli:2016xae,Cicoli:2017axo,Cicoli:2018tcq,Cicoli:2016chb, Cicoli:2020bao, Burgess:2016owb, Cicoli:2011ct}. We discuss our results and present our conclusions in Sec. \ref{Conclusions}. Three appendices are devoted to provide technical details: App. \ref{AppCanNorm} gives the details of the axion canonical normalisation; App. \ref{abun} provides a few benchmark points which reproduce the observed fuzzy DM abundance for ultra-light stringy ALPs; App. \ref{OtherDistributions} shows the distribution of additional quantities relevant for phenomenology, like moduli masses and the reheating temperature from moduli decay, which also feature a logarithmic distribution in the flux landscape.

\section{Axions and moduli stabilisation}
\label{sec:LVS}

As explained in Sec. \ref{Intro}, axions can be light (i.e. much lighter than the gravitino and the soft terms) only if supersymmetry is broken and the corresponding saxions are fixed at perturbative level. Moreover models with a large number of K\"ahler moduli require a huge CY volume to keep control over the $\alpha'$ expansion. These two considerations single out type IIB LVS models as the best framework to study the interplay between axion physics and moduli stabilisation. 

We shall now describe moduli stabilisation for a toy-model which can feature up to 3 light axions. This model is, at the same time, simple enough to perform moduli stabilisation in full detail, and rich enough to be a good representative of a more generic situation. In fact, it has 1 axion which becomes as heavy as the gravitino because of non-perturbative stabilisation, 1 ultra-light bulk axion which plays the role of an ALP, and 2 QCD axion candidates arising from the reduction of $C_4$ over a bulk or a local 4-cycle.

\subsection{The geometry}
\label{section2.1}

The total number of K\"ahler moduli is $h^{1,1}(X)=4$ and the CY $X$ features a K3 or $T^4$ divisor $D_1$ fibred over a $\mathbb{P}^1$ base contained in a second divisor $D_2$, and two additional rigid divisors $D_3$ and $D_4$ with only self-intersections. The K\"ahler form can be expanded in a basis of $(1,1)$-forms as $J=t_1 \hat{D}_1+t_2\hat{D}_2-t_3\hat{D_3}-t_4\hat{D}_4$ where the $t_i$ are 2-cycle volumes and the negative signs have been chosen to ensure that all 2-cycle volumes are positive (in particular those dual to rigid divisors). The only non-vanishing intersection numbers are $k_{122}$, $k_{333}$ and $k_{444}$. Explicit examples with these properties can be found in \cite{Cicoli:2011it, Cicoli:2016xae, Cicoli:2017axo}. Thus the CY volume form looks like:
\be
\vo = \frac16\int_X J\wedge J\wedge J = \frac12 k_{122} \,t_1 t_2^2  -\frac16 k_{333}\, t_3^3-\frac16 k_{444}\,  t_4^3\,.
\ee
The 4-cycle moduli $\tau_i=\frac12 \int_X \hat{D}_i\wedge J\wedge J$ become:
\be
\tau_1 = \frac12 k_{122}t_2^2\,,\qquad \tau_2 = k_{122} t_1 t_2\,, \qquad
\tau_3 = \frac12 k_{333}\,t_3^2\,,\qquad
\tau_4 = \frac12 k_{444}\,t_4^2\,.
\ee
These relations can be inverted and $\vo$ can be written in terms of 4-cycle moduli as:
\be
\vo = \alpha\left(\sqrt{\tau_1}\tau_2-\gamma_3\tau_3^{3/2}-\gamma_4\tau_4^{3/2}\right),
\label{voform}
\ee
where $\alpha = \frac{1}{\sqrt{2 k_{122}}}$, $\gamma_3=\frac23\sqrt{\frac{k_{122}}{k_{333}}}$ and $\gamma_4=\frac23 \sqrt{\frac{k_{122}}{k_{444}}}$.

Before dwelling on the details of moduli stabilisation, let us outline the main features of this representative model. We assume that the SM is built on stacks of D7-branes wrapping a 4-cycle in the geometric regime. As typical of LVS models, the internal volume is stabilised at exponentially large values. On the other hand, the 2 blow-up modes $\tau_3$ and $\tau_4$ are fixed at small values, and so the volume can be approximated as $\vo\simeq \alpha \sqrt{\tau_1}\tau_2$. Given that $\vo$ is controlled by 2 moduli, we can consider 2 different regimes in moduli space: 
\ben
\item \textbf{Isotropic limit with SM on a local cycle:} In this case $\tau_1\sim \tau_2\gg \tau_3\sim \tau_4$. Both $\tau_1$ and $\tau_2$ are exponentially large, and so none of them can be wrapped by the SM D7-stack since the corresponding gauge coupling $\alpha_\SM^{-1}=\tau_i$, $i=1,2$, would be hyper-weak. Hence the SM lives on a D7-stack wrapping the local divisor $D_3$. $\tau_4$ and $\theta_4$ are fixed by instantons which make both of them as heavy as the gravitino. $\tau_1$ and $\tau_2$ are fixed by a combination of $\alpha'$ and $g_s$ effects, and so the corresponding axions $\theta_1$ and $\theta_2$ are ultra-light ALPs. $\tau_3$ is stabilised by a combination of D-terms, F-terms of matter fields and string loops. The associated axion $\theta_3$ plays the role of the QCD axion with a decay constant of order $M_s$ which is around the intermediate scale for TeV-scale soft terms.

\item \textbf{Anisotropic limit with SM on a bulk cycle:} In this case $\tau_2\gg \tau_1\sim \tau_3\sim \tau_4$. $\tau_1$ and $\tau_2$ are again frozen by perturbative corrections to the K\"ahler potential, and $\tau_3$ by non-perturbative contributions to the superpotential. Contrary to the previous case, $\tau_3$ is instead stabilised by non-perturbative physics. Given that $\tau_1$ is hierarchically smaller than $\tau_2$, the underlying CY has an anisotropic shape with 2 extra dimensions much larger than the other 4. Thus the SM can live on the bulk divisor $D_1$. $\theta_1$ becomes the QCD axion with a decay constant set by the Kaluza-Klein scale associated to the fibre divisor $D_1$ which turns out to be of order the GUT scale. The mass of $\theta_3$ and $\theta_4$ is around $m_{3/2}$, whereas $\theta_1$ plays again the role of an ultra-light ALP.
\een

\subsection{Moduli stabilisation: leading results}
\label{ModStabLead}

The model-independent closed string moduli involve the axion dilaton $S=e^{-\phi}+iC_0$, $h^{1,2}(X)$ complex structure moduli $U_a$, and $h^{1,1}(X)=4$ K\"ahler moduli $T_i= \tau_i+{\rm i}\,\theta_i$. Dimensional reduction yields the following tree-level K\"ahler potential:\footnote{Here and in the following we set $M_p=1$ but we will reinsert the correct powers of $M_p$ in the main results.}
\be
K_{\rm tree}=-2 \ln \vo -\ln\left(S+\bar{S}\right)+\ln\left(-i\int_{X}\Omega \wedge \bar{\Omega}\right).
\label{Ktree}
\ee
$S$ and the $U$-moduli are fixed at tree-level by turning on the 3-form flux $G_3=F_3+iSH_3$ which generates the superpotential:
\be
W_{\rm tree} = \int_X G_3 \wedge \Omega(U)\,,
\label{Wtree}
\ee
where $\Omega(U)$ is the $U$-dependent holomorphic $(3,0)$-form of $X$. The complex structure moduli and the axio-dilaton develop a mass of order $m_{3/2}$, and so the associated axions are too heavy to be relevant for low-energy phenomenology.

Because of the no-scale cancellation, the $T$-moduli remain flat at semi-classical order. These directions are lifted by including non-perturbative corrections to (\ref{Wtree}) and perturbative corrections to (\ref{Ktree}). Focusing just on the K\"ahler sector and including only the leading order $\alpha'$ effects and instanton contributions, $K$ and $W$ become:
\be
K=-2 \ln\left(\vo+\frac{\xi}{2g_s^{3/2}}\right),
\qquad
W = W_0+ A_4\, e^{-\mathfrak{a}_4 T_4}\,,
\label{KandWcorr}
\ee
where $\xi = -\frac{\chi(X)\zeta(3)}{2(2\pi)^3}$ with $\chi(X)$ the Euler number of $X$ and $\zeta(3)\simeq 1.2$, $W_0$ is the vacuum expectation value of (\ref{Wtree}), $A_4\sim\mc{O}(1)$ and $\mathfrak{a}_4=2\pi$ for an ED3 wrapping $D_4$, while $\mathfrak{a}_4=2\pi/\mathfrak{n}_4$ for gaugino condensation on a stack of $\mathfrak{n}_4$ D7-branes on $D_4$. 

Plugging (\ref{KandWcorr}) into the standard form of the 4D $N=1$ supergravity F-term scalar potential, we end up with (up to an overall $S$ and $U$-dependent factor):
\be
V = \frac{8}{3\alpha^2} \mathfrak{a}_4^2 A_4^2 \frac{\sqrt{\tau_4}}{\vo} \,e^{-2\mathfrak{a}_4\tau_4}
+ 4 \mathfrak{a}_4 A_4 \tau_4 \cos(\mathfrak{a}_4 \theta_4) \frac{W_0}{\vo^2} e^{-\mathfrak{a}_4 \tau_4}
+\frac{3\xi}{4 g_s^{3/2}}\frac{W_0^2}{\vo^3}\,.
\label{LVSpot}
\ee
Minimising (\ref{LVSpot}) with respect to the 3 moduli $\vo$, $\tau_4$ and $\theta_4$ results in:
\be
\langle \vo \rangle = \frac{3\alpha}{4 \mathfrak{a}_4 A_4}\sqrt{\langle \tau_4 \rangle} W_0\, e^{\mathfrak{a}_4 \langle \tau_4 \rangle},  \qquad \langle \tau_4 \rangle = \frac{1}{g_s}\left(\frac{\xi}{2\alpha} \right)^{2/3},
\qquad
\langle\theta_4\rangle = (2k+1)\frac{\pi}{\mathfrak{a}_4}\,\,k\in\mathbb{Z}\,.
\label{swiss_minimum}
\ee
This vacuum is AdS but there exist several mechanisms to uplift it to a dS solution \cite{Cicoli:2012fh, Cicoli:2015ylx,Gallego:2017dvd,Crino:2020qwk}. The order of magnitude of the induced moduli masses is:
\be
m_{\tau_4}\simeq m_{\theta_4}\simeq m_{3/2}=\sqrt{\frac{g_s}{2\pi}}\frac{W_0 M_p}{\vo},\qquad m_\vo\simeq m_{3/2}\sqrt{\frac{m_{3/2}}{M_p}}\,,
\label{LeadMass}
\ee
showing that the axion $\theta_4$ becomes too heavy to be relevant for low-energy phenomenology since $m_{3/2}$ sets also the order of magnitude of the soft terms, $M_{\rm soft}\simeq m_{3/2}$, which cannot be below the TeV-scale. At this level of approximation all the other K\"ahler moduli are still flat.

\subsection{Moduli stabilisation: subleading results and axion physics}
\label{LeadingResults}

Let us now describe the stabilisation of the remaining K\"ahler moduli by including additional contributions to $K$ and $W$ which are subdominant with respect to those considered in Sec. \ref{ModStabLead}. The small parameters controlling the $g_s$ and $\alpha'$
expansions are respectively $e^\phi\ll 1$ and $\vo^{-1/3}\ll 1$. We shall consider the isotropic and anisotropic limits separately.

\subsubsection{Isotropic limit with SM on a local cycle}
\label{IsotropicLimit}

In this case the SM lives on D7-branes wrapped around the `small' rigid divisor $D_3$. Because of the well-known tension between chirality and non-perturbative effects \cite{Blumenhagen:2007sm}, $\tau_3$ cannot be stabilised by instantons, and so $\theta_3$, contrary to $\theta_4$, remains light and can play the role of the QCD axion. Let us see this important issue in detail.

\subsubsection*{Moduli stabilisation}

The total world-volume fluxes on the SM D7-stack and an ED3 instanton (similar considerations apply to gaugino condensation) on $D_3$ look like:
\be
\mc{F}_\SM = f_\SM \hat{D}_3 + \frac12 \hat{D}_3 -B\,,\qquad \mc{F}_\ED3 = \frac12 \hat{D}_3-B\,,
\ee
where $f_\SM \in \mathbb{Z}$ and the half-integer contributions are due to Freed-Witten anomaly cancellation on non-spin divisors \cite{Minasian:1997mm, Freed:1999vc}. In order to obtain an $O(1)$ instanton which contributes to $W$, the $B$-field has to chosen as $B=\frac12 \hat{D}_3$ so that $\mc{F}_\ED3 =0$. This, in turn, gives:
\be
\mc{F}_\SM = f_\SM \hat{D}_3\,,\qquad \mc{F}_\ED3 = 0\,.
\label{GaugeFlux}
\ee
The number of chiral intersections between the ED3 and the SM D7-stack is then given by:
\be
I_{\SM-\ED3} = \int_X \left(\mc{F}_\SM-\mc{F}_\ED3\right) \wedge \hat{D}_3\wedge \hat{D}_3  = k_{333} f_\SM\,.
\ee
These zero modes can kill the ED3 contribution to $W$ if they develop vanishing vacuum expectation values, as expected for visible sector fields in order not to break any of the SM gauge symmetries at high energies. In fact, the gauge flux $\mc{F}_\SM$ in (\ref{GaugeFlux}) induces the following $U(1)$-charge for the modulus $T_3$:
\be
q_{T_3} = \int_X \mc{F}_\SM \wedge \hat{D}_3\wedge \hat{D}_3 = k_{333} f_\SM\,,
\ee
and $W$ has to be gauge invariant. This implies that the prefactor of the non-perturbative $W$ has also to depend on charged matter fields. Considering for simplicity just a single open string field $\phi$, the relevant $U(1)$ transformations are:
\be
\delta \phi = {\rm i} q_\phi \phi\,,\qquad \delta T_3 = {\rm i}\,\frac{q_{T_3}}{2\pi}\,.
\ee
Thus the non-perturbative superpotential (including the possibility of gaugino condensation):
\be
W_\ED3 = A_3\,e^{-\mathfrak{a}_3 T_3}\qquad\text{with}\qquad A_3 = A \phi^n\quad\text{and}\quad \mathfrak{a}_3=\frac{2\pi}{\mathfrak{n}_3}\,,
\label{WED3}
\ee
transforms under the anomalous $U(1)$ as:
\be
\delta W_\ED3 = W_\ED3 \left(n\,\frac{\delta\phi}{\phi}-\frac{2\pi}{\mathfrak{n}_3} \delta T_3\right) = {\rm i} W_\ED3 \left(n\,q_\phi-\frac{q_{T_3}}{\mathfrak{n}_3}\right)\,,
\ee
implying that $W_\ED3$ can be gauge invariant if $n = q_{T_3}/(\mathfrak{n}_3 q_\phi)$. As can be clearly seen from (\ref{WED3}), $A_3 = 0$ if $\langle \phi\rangle=0$ (for $n>0$). 

This problem comes along with the following correlated issue. A non-zero gauge flux on the D7-stack generates also a moduli-dependent Fayet-Iliopoulos term of the form \cite{Jockers:2005zy, Haack:2006cy}:
\be
\xi_\SM = \frac{1}{4\pi\vo}\int_X J\wedge \mc{F}_\SM \wedge \hat{D}_3 = -\frac{q_{T_3}}{4\pi}\frac{t_3}{\vo} = - \frac{f_\SM\sqrt{2 k_{333}}}{4\pi}\frac{\sqrt{\tau_3}}{\vo}\,.
\label{FIterm}
\ee
If $\langle\phi\rangle = 0$, a vanishing D-term potential requires $\xi_\SM=0$ which, in turn, implies $\tau_3\to 0$, causing the collapse of the divisor $D_3$ to a singularity. 
This shrinking can be avoided in 2 ways: ($i$) by considering a slightly different geometry where the 2 rigid divisors $D_3$ and $D_4$ intersect each other so that $\xi_\SM=0$ would just fix $\tau_3$ in terms of $\tau_4$; ($ii$) by considering the case where $\phi$ is a SM gauge singlet (like a right handed sneutrino) which can develop a non-zero vacuum expectation value by D-term cancellation. 

In what follows we shall focus on the option ($ii$) since in the case ($i$) the anomalous $U(1)$ would become massive by eating up a combination of the $\theta_3$ and $\theta_4$, leaving no light closed string axions to behave as the QCD axion.

The D-term potential reads (taking, without loss of generality, $\phi$ as a canonically normalised field):
\be
V_D = \frac{g_\SM^2}{2}\left(q_\phi |\phi|^2 + \xi_\SM\right)^2\,.
\ee
A vanishing D-term potential then fixes the open string field at:
\be
\langle |\phi|^2\rangle = \frac{n}{4\pi}\frac{t_3}{\vo} = c\, \frac{\sqrt{\tau_3}}{\vo}\qquad\text{with}\quad c = \frac{n}{4\pi}\sqrt{\frac{2}{k_{333}}}\,.
\label{phiVEV}
\ee
The anomalous $U(1)$ becomes massive by eating up a combination of $\theta_3$ and the phase $\theta_\phi$ of $\phi =|\phi|\,e^{{\rm i} \theta_\phi}$. Its mass is given by:
\be
M_{U(1)}^2\simeq g_\SM^2 M_p^2 \left(f_{\rm op}^2+f_{\rm cl}^2\right),
\label{MU(1)}
\ee
where:
\be
f_{\rm op}^2 = \langle |\phi|^2 \rangle = c\,\frac{\sqrt{\tau_3}}{\vo}\,,
\label{fop}
\ee
is the decay constant of the open string axion $\theta_\phi$, while $f_{\rm cl}$ is the decay constant of the closed string axion $\theta_3$. This last quantity can be derived from the kinetic terms:
\be
\mc{L}_{\rm kin} \supset \frac14\frac{\partial^2 K}{\partial \tau_3^2}\partial_\mu \theta_3\partial^\mu\theta_3 = \frac12 \partial_\mu a\partial^\mu a\,,
\ee
where $a \simeq \theta_3 f_{\rm cl}$ is the canonically normalised axion, implying:
\be
f_{\rm cl}^2 = \frac12 \frac{\partial^2 K}{\partial \tau_3^2} = \frac18\sqrt{\frac{ k_{122}}{k_{333}}}\frac{1}{\vo \sqrt{\tau_3}}\,.
\label{fcl}
\ee
Comparing (\ref{fop}) with (\ref{fcl}), it is easy to see that $f_{\rm op}\gg f_{\rm cl}$ for $\tau_3\gg 1$, signaling that the combination of $\theta_3$ and $\theta_\phi$ eaten up by the anomalous $U(1)$ is mostly given by the open string axion $\theta_\phi$ since the largest contribution to $M_{U(1)}$ in (\ref{MU(1)}) comes from $f_{\rm op}$ \cite{Allahverdi:2014ppa}. Thus $\theta_3$ survives in the low-energy theory and can play the role of the QCD axion $a$. The corresponding saxion $\tau_3$ develops a potential via 2 effects:
\ben
\item The F-term potential of the matter field $\phi$ generated by supersymmetry breaking effects, after writing $|\phi|$ in terms of $\tau_3$ using (\ref{phiVEV}):
\be
V_{\rm matter} = m_{3/2}^2 |\phi|^2 = c\, \frac{W_0^2 \sqrt{\tau_3}}{\vo^3}\,.
\ee

\item The potential generated by string loop corrections to the K\"ahler potential due to the exchange of Kaluza-Klein modes between the D7-stack wrapped around $D_3$ and O7-planes or D3-branes \cite{Berg:2005ja, Berg:2007wt}:
\be
V_{\rm loop} = c_{\rm loop}\, \frac{W_0^2}{\vo^3\sqrt{\tau_3}}\,,
\ee
where $c_{\rm loop}$ is expected to be an $\mc{O}(1-10)$ coefficient which depends on the $U$-moduli. 
\een
The potential $V_{\rm matter}+V_{\rm loop}$ admits a minimum at $\langle\tau_3\rangle = c_{\rm loop}/c \sim \mc{O}(10)$ which reproduces the correct order of magnitude of the SM gauge coupling $g_\SM^{-2}\simeq \langle\tau_3\rangle$. It can be easily checked that the saxion $\tau_3$ develops a mass of order $m_{3/2}$ similarly to $\tau_4$ and $\theta_4$. 

Notice that $T_3$-dependent instanton corrections to $W$ as in (\ref{WED3}) would generate a potential of the form (using (\ref{phiVEV}) and setting $\theta_\phi=0$ and $\mathfrak{n}_3=1$):
\be
V =  \frac{32 A c \pi^2}{3\alpha^2} \frac{\tau_3^{(1+n)/2}}{\vo^{1+n}} \,e^{-4\pi\tau_3}
+ 8\pi A c^{n/2} \tau_3^{1+n/4} \cos(2\pi \theta_3)\, \frac{W_0}{\vo^{2+n/2}}\, e^{-2\pi \tau_3}\,.
\label{Vnp4}
\ee
For $n=q_{T_3}/q_\phi \geq 2$ and $2\pi\tau_3\gg 1$ (i.e. the limit where higher-order instanton corrections can be neglected), the potential (\ref{Vnp4}) is exponentially suppressed with respect to $V_{\rm matter}+V_{\rm loop}$, and so it produces just a tiny shift of the minimum for $\tau_3$. On the other hand, it would generate a mass for $\theta_3$ of order: 
\be
m_{\theta_3}\simeq \sqrt{K^{-1}_{33} V_{\theta_3\theta_3}} \sim  \left(\frac{m_{3/2}}{M_p}\right)^{\frac{n-2}{4}}\,m_{3/2} \,e^{-\pi \tau_3}\,,
\ee
which for $n\geq 2$, $m_{3/2}\lesssim \mc{O}(10^{10})$ GeV and $\tau_3=\alpha_\SM^{-1}\simeq 25$ is always subdominant with respect to the contribution from QCD instantons $m_a \simeq \Lambda_\QCD^2/f_a$ for any possible $f_a\lesssim M_p$. This guarantees that $\theta_3$ is a good QCD axion candidate.

The only saxion which remains to be fixed is the fibre modulus $\tau_1$. This field develops a potential via string loop corrections which experience an `extended no-scale cancellation' that suppresses them with respect to the leading $\alpha'$ correction \cite{Cicoli:2007xp}. The resulting scalar potential for $\tau_1$ reads \cite{Cicoli:2008gp}:\footnote{We neglected loop corrections suppressed by additional powers of $g_s\ll 1$.}
\be
V_{g_s}=\left(g_s^2 \frac{A}{\tau_1^2}-\frac{B}{\vo\sqrt{\tau_1}}\right)\frac{W_0^2}{\vo^2}\,,
\label{tau1pot}
\ee
where $A$ and $B$ are flux-dependent parameters. The minimum of (\ref{tau1pot}) is located at:
\be
\langle \tau_1 \rangle  = \lambda g_s^{4/3} \langle \vo \rangle^{2/3}= \alpha \lambda^{3/2} g_s^2 \langle \tau_2 \rangle\,,\qquad \lambda \equiv \left(\frac{4A}{B} \right)^{2/3}\,.
\label{tau1min}
\ee
This result has 3 important implications:
\ben
\item For $\alpha\simeq\lambda\simeq\mc{O}(1)$ and $g_s \simeq\mc{O}(0.1)$, $\tau_1$ is roughly of the same order as $\tau_2$, implying that the CY volume is isotropic. Without loss of generality, in what follows we shall consider $\alpha\lambda^{3/2}g_s^2=1$, i.e. $\langle\tau_1\rangle=\langle\tau_2\rangle$.

\item The scalar potential (\ref{tau1pot}) scales as $\vo^{-10/3}$, and so for $\vo \gg 1$ it is indeed suppressed with respect to the leading order LVS potential (\ref{LVSpot}) which scales as $\vo^{-3}$.

\item For $\vo \gg 1$ the SM cannot live either on $\tau_1$ or on $\tau_2$ since the resulting gauge coupling would be too small. Hence the SM has to be supported by $\tau_3$. 
\een
Notice that the axions $\theta_1$ and $\theta_2$ are lifted only by tiny non-perturbative corrections to the superpotential of the form: 
\be
W\supset A_1\, e^{-\mathfrak{a}_1 T_1} + A_2\,e^{-\mathfrak{a}_2 T_2}\,.
\ee
with $A_1\simeq A_2\simeq\mc{O}(1)$, and $\mathfrak{a}_i = 2\pi/\mathfrak{n}_i$ for $i=1,2$. Given that $\tau_1\simeq\tau_2\gg 1$, these effects would make $\theta_1$ and $\theta_2$ 2 ultra-light, i.e. almost massless, ALPs. 

\subsubsection*{Mass spectrum and decay constants}

The mass spectrum of the 3 moduli fixed at leading order, $\vo$, $\tau_4$ and $\theta_4$ has been given in (\ref{LeadMass}). The mass of the remaining moduli turns out to be:
\bea
m_{\tau_3}&\simeq& m_{3/2}\,,\qquad
m_{\tau_1}\simeq m_{3/2}\left(\frac{m_{3/2}}{M_p}\right)^{2/3}\,, \nn \\
m_{\theta_3}&\equiv& m_a \simeq \frac{\Lambda_\QCD^2}{f_{\theta_3}}\,,
\qquad m_{\theta_1} \simeq M_p\,e^{-\pi\tau_1/\mathfrak{n}_1}
\,,\qquad m_{\theta_2} \simeq M_p\,e^{-\pi\tau_2/\mathfrak{n}_2}\,.
\label{m_iso}
\eea
Notice that $\alpha'$ effects are under control when $\vo\simeq \tau_1^{3/2}\simeq
\tau_2^{3/2} \gtrsim 10^3$ since the corresponding expansion parameter is $\vo^{-1/3}\lesssim 0.1$. In this regime the 2 ALPs $\theta_1$ and $\theta_2$ are almost massless, $m_{\theta_1}\sim 0$ and $m_{\theta_2}\sim 0$, since their mass would turn out to be smaller than the present value of the Hubble constant, $H_0\simeq 10^{-33}$ eV, for $\mathfrak{n_1}=\mathfrak{n_2}=1$. Larger values of $\mathfrak{n_1}$ and $\mathfrak{n_2}$ can however raise $m_{\theta_1}$ and $m_{\theta_2}$ above $H_0$, with interesting application to fuzzy DM \cite{Hui:2016ltb}. In what follows we shall therefore consider $\theta_1$ and $\theta_2$ as ultra-light.

The decay constants of the QCD axion $\theta_3$ and the 2 ALPs $\theta_1$ and $\theta_2$ can be derived from canonical normalisation and take the generic form (see \cite{Cicoli:2012sz} and App. \ref{fdef}):
\be
f_{\theta_i}\equiv \left(\frac{\mathfrak{n}_i}{2\pi}\right)\sqrt{2\lambda_i}\,M_p\,,
\ee
where $\lambda_i$ is the $i$-th eigenvalue of the K\"ahler metric and $\mathfrak{n}_i$ determines the periodicity of the cosine potential which enjoys a discrete shift symmetry (with $\mathfrak{n}_3=1$ for the QCD axion). The details of canonical normalisation for this explicit example are provided in App. \ref{AppCanNorm}. The eigenvalues of the K\"ahler metric (\ref{fib_Kahler_metr}) are given by $\frac{1}{2\tau_2^2}$, $\frac{1}{4\tau_1^2}$ and $\frac{3\alpha\gamma_3}{8}\frac{1}{\vo\sqrt{\tau_3}}$, and so the decay constants become:
\be
f_{\theta_3}\equiv f_a =\frac{c_3}{\langle\tau_3\rangle^{1/4}}\frac{M_p}{\sqrt{\langle\vo\rangle}}\,,\quad f_{\theta_1}=c_1\frac{M_p}{\langle\tau_1\rangle} = c_1\alpha^{2/3}\frac{M_p}{\langle\vo\rangle^{2/3}}\,,\quad f_{\theta_2}=c_2\frac{M_p}{\langle\tau_2\rangle}=c_2\alpha^{2/3}\frac{M_p}{\langle\vo\rangle^{2/3}}\,,
\label{f_iso}
\ee
where the $c_i$'s are moduli-independent coefficients:
\be
c_3 = \frac{\sqrt{3\alpha\gamma_3}}{4\pi}\,, \qquad
c_1 =\frac{\mathfrak{n}_1}{2\pi\sqrt{2}}\,, \qquad
c_2 = \frac{\mathfrak{n}_2}{2\pi}\,.
\label{cs}
\ee
Notice that the QCD axion decay constant scales as the string scale, $f_a \simeq M_s\simeq M_p/\sqrt{\vo}$, while the decay constants of the 2 ultra-light ALPs behave as the Kaluza-Klein scale, $f_{\theta_1}\simeq f_{\theta_2}\simeq M_\KK\simeq M_p/\vo^{2/3}$. 

The order of magnitude of all these mass scales is set by the overall volume $\vo$. An interesting regime in moduli space is the one where $W_0\sim \mc{O}(1-10)$ and $\vo\sim\mc{O}(10^{14-15})$ which leads to TeV-scale supersymmetry, $M_{\rm soft}\sim m_{3/2}\sim\mc{O}(1-10)$ TeV, and a QCD axion decay constant at intermediate scales, $f_a\sim M_s \sim \mc{O}(10^{10}-10^{11})$ GeV. Smaller values of the CY volume, like $\vo\sim\mc{O}(10^3-10^4)$, would push $M_{\rm soft}$ to intermediate scales and $f_a$ around the GUT scale.

\subsubsection*{Axion couplings to gauge bosons}

Other quantities which are relevant for phenomenology are the couplings of the axions to the gauge fields of the visible and hidden sectors. We shall focus just on the couplings of the QCD axion and the 2 ultra-light ALPs which we will express in terms of the corresponding canonically normalised fields $a_3\equiv a$, $a_1$ and $a_2$ (see App. \ref{AppCanNorm} for the details of canonical normalisation). The visible sector lives on $D_3$ while the hidden sector involves 2 intersecting stacks of D7-branes wrapped around $D_1$ and $D_2$. Knowing that the gauge kinetic function of each sector is given by the corresponding unnormalised K\"ahler modulus, and denoting the field strengths of the canonically normalised gauge bosons respectively as $F_{\rm vis}$, $F_1$ and $F_2$, we obtain: 
\be
\mc{L}_{\rm ax-gauge} =  \frac{a}{f_a}\left[\frac{\lambda_1}{\langle\tau_3\rangle}\, \tilde F_{\rm vis} F_{\rm vis} + \frac{\sqrt{\langle\tau_3\rangle}}{\vov}\left(\lambda_2\tilde{F}_1 F_1 + \lambda_3\tilde{F}_2 F_2\right) \right] 
+ \lambda_4\frac{a_1}{M_p} \tilde{F}_1 F_1 + \lambda_5\frac{a_2}{M_p} \tilde{F}_2 F_2,
\label{fib_canon_couplings_iso}
\ee
where the $\lambda_i$'s are numerical $\mc{O}(1)$ coefficients. Notice that the QCD axion $a$ has a stronger than Planckian coupling to the visible gauge bosons while its coupling to hidden sector degrees of freedom is very suppressed. On the other hand, the 2 ALPs have a standard $\mc{O}(1/M_p)$ coupling to hidden gauge bosons but they are decoupled from the visible sector. These results are due to the combination of two effects: ($i$) the visible sector lives on a shrinkable del Pezzo $D_3$ which has no intersection with the bulk divisors $D_1$ and $D_2$; ($ii$) the axions $\theta_1$ and $\theta_2$ are in practice massless.

\subsubsection{Anisotropic limit with SM on a bulk cycle}
\label{AnisSM}

In LVS scenarios the SM can be realised on a stack of D7-branes wrapped around a bulk cycle only if the underlying geometry has an anisotropic shape. In this case the overall volume can be exponentially large in agreement with a SM gauge coupling which is not too small. 

\subsubsection*{Moduli stabilisation}

We focus on the case where the SM lives on the K3 or $T^4$ fibre $D_1$. Hence the visible sector gauge coupling is given by $\alpha_\SM^{-1} = \tau_1\simeq\mc{O}(10$-$100)$. Given that $\vo\simeq\alpha\sqrt{\tau_1}\tau_2$ is exponentially large, the internal geometry needs to be anisotropic with 2 extra dimensions much larger than the other 4. This can be achieved via the following moduli stabilisation procedure:
\bi
\item $\vo$, $\tau_4$ and $\theta_4$ are stabilised as in  (\ref{swiss_minimum}) and the CY volume becomes exponentially large in string units.

\item Given that $D_3$ is not wrapped by the SM D7-stack, the non-perturbative superpotential (\ref{WED3}) is not suppressed anymore due to chiral intersections with visible sector states. Hence the freezing of $\tau_3$ and $\theta_3$ is completely similar to the stabilisation of $\tau_4$ and $\theta_4$. Contrary to the isotropic scenario, in this case $\theta_3$ acquires a mass of order $m_{3/2}$ and plays no role for low-energy physics. 

\item The fibre divisor $\tau_1$ is stabilised by string loop corrections as in (\ref{tau1min}) but with $\alpha \lambda^{3/2}=1$ and $g_s\ll 1$.\footnote{To be more precise, we envisage a situation similar to the explicit CY cases discussed in \cite{Cicoli:2017axo} where the volume (neglecting blow-up modes) is $\vo \simeq \sqrt{\tau_1\tau_2\tilde\tau_2}$. Due to the intersection between $\tau_2$ and $\tilde\tau_2$, the Fayet-Iliopoulos term induced by gauge fluxes fixes $\tau_2\propto\tilde\tau_2$ for vanishing VEVs of open string fields. An appropriate combination of closed string axions is eaten up by the corresponding anomalous $U(1)$. Substituting $\tau_2\propto\tilde\tau_2$ in $\vo$, one obtains effectively the same expression that we are considering: $\vo \simeq \sqrt{\tau_1}\tau_2$.} This results in the following hierarchy:
\be
\langle\tau_1\rangle = g_s^2\,\langle\tau_2\rangle \ll \langle\tau_2\rangle\,.
\label{tau1minNew}
\ee

\item $T_1$-dependent non-perturbative corrections to $W$ would be very suppressed due to: ($i$) the presence of chiral intersections as for $T_3$-dependent instantons in the isotropic case; ($ii$) the fact that $D_1$ is a non-rigid cycle with extra fermionic zero modes which tend to kill instanton contributions. Therefore the closed string axion $\theta_1$ is a perfect QCD axion candidate which becomes massive via standard QCD instantons.

\item The remaining closed string axion $\theta_2$ is an almost massless ALP which develops a tiny mass via non-perturbative corrections to $W$ which are exponentially suppressed in terms of the large 4-cycle $\tau_2$.
\ei

\subsubsection*{Mass spectrum and decay constants}

The mass of $\vo$, $\tau_4$ and $\theta_4$ is again given by (\ref{LeadMass}). The mass spectrum of the other moduli instead reads:
\bea
m_{\tau_3}&\simeq& m_{\theta_3}\simeq m_{3/2}\,,\qquad
m_{\tau_1}\lesssim m_{3/2}\sqrt{\frac{m_{3/2}}{M_p}}\,, \nn \\
m_{\theta_1}&\equiv& m_a \simeq \frac{\Lambda_\QCD^2}{f_{\theta_1}}\,,
\qquad m_{\theta_2} \simeq M_p\,e^{-\pi\tau_2/\mathfrak{n}_2}\,.
\label{mass_aniso}
\eea
The decay constants of the QCD axion $\theta_1$ and the ALP $\theta_2$ now become:
\be
f_{\theta_1}\equiv f_a =c_1\frac{M_p}{\langle\tau_1\rangle} = c_1\alpha_\SM M_p\,,\qquad f_{\theta_2}  =c_2\frac{M_p}{\langle\tau_2\rangle}=c_2\alpha_\SM g_s^2 M_p\,,
\label{f_aniso}
\ee
where $c_1$ and $c_2$ are again given by (\ref{cs}) with $\mathfrak{n}_1=1$ for the QCD axion $\theta_1$. Notice that the QCD axion decay constant is proportional to the visible sector gauge coupling since $\alpha_\SM^{-1} =  \langle\tau_1\rangle$ and so $\alpha_\SM^{-1}\sim\mc{O}(10-100)$ implies a GUT-scale decay constant $f_a\simeq M_\GUT$. Contrary to the isotropic scenario where the SM was supported on a local cycle and the QCD axion decay constant could take different values from $M_\GUT$ to intermediate scales depending on the value of $\vo$, in this case the QCD axion decay constant is fixed at $M_\GUT$ by the requirement of reproducing the observed value of the SM gauge coupling.

Moreover, for $g_s\lesssim 0.1$ and $\langle\tau_1\rangle\gtrsim 10$, (\ref{tau1minNew}) yields $\langle\tau_2\rangle\gtrsim 10^3$ which, in turn, implies that the ALP $\theta_2$ is ultra-light, i.e. $m_{\theta_2}\sim 0$. The decay constant of this ALP is set by the Kaluza-Klein scale of the effective 6D theory since:
\be
M_\KK^\sixD \simeq \frac{M_s}{\sqrt{t_1}} \simeq \frac{M_p}{\alpha \tau_2}\simeq f_{\theta_2}\,.
\ee
Interestingly, this scale is one order of magnitude above the gravitino mass since:
\be
m_{3/2}\simeq\frac{M_p}{\vo}\simeq \sqrt{\alpha_\SM}\, M_\KK^\sixD \sim \mc{O}(0.1)\, M_\KK^\sixD\,.
\label{6Dm32}
\ee
The decay constant $f_{\theta_2}$ can take different values depending on the order of magnitude of $\tau_2$. Varying $\tau_2$ corresponds to varying $\vo$ which is mainly controlled by $g_s$, as can be seen from (\ref{swiss_minimum}). The string coupling affects also the relation (\ref{tau1minNew}) where however $\tau_1$ has to remain fixed to get the right SM gauge coupling. This can be achieved by varying $W_0$ as well. Hence (\ref{swiss_minimum}), (\ref{tau1minNew}) and $\tau_1= \alpha_\SM^{-1} \sim \mc{O}(10-100)$ imply an interesting relation between $W_0$ and $g_s$ (ignoring $\mc{O}(1)$ numerical factors):
\be
W_0 \sim \left(\frac{\alpha_\SM^{-1}}{g_s}\right)^{3/2} e^{-\frac{\mathfrak{a}_4}{g_s}}\sim \mc{O}(10^3)\,g_s^{-3/2}\,e^{-1/g_s}\,.
\label{W0gsRel}
\ee
Thus this class of constructions can reproduce the right visible sector gauge coupling only for flux vacua which satisfy the relation (\ref{W0gsRel}). This implies that larger scales are more natural since $f_{\theta_2}\simeq 10^{14}$ GeV can be obtained for $g_s\simeq\mc{O}(0.1)$ and $W_0\simeq\mc{O}(1)$, but $f_{\theta_2}\simeq 10^{12}$ GeV needs $g_s\simeq\mc{O}(0.01)$ that would require a severe tuning of $W_0$ down to values of order $W_0\simeq\mc{O}(10^{-38})$. We conclude that this scenario naturally predicts a QCD axion decay constant around the GUT scale, an almost massless ALP with decay constant around $10^{14}$ GeV and supersymmetry at intermediate scales.

\subsubsection*{Axion couplings to gauge bosons}

Let us now focus on the coupling of the canonically normalised QCD axion $a_1\equiv a$ and ALP $a_2$ to gauge bosons belonging to the visible sector on $D_1$ and hidden sectors on $D_2$ and $D_3$. In fact, SM particles are not charged under the gauge symmetries of the D7-stack wrapping $D_3$ since $D_3$ does not intersect with $D_1$. Similar considerations apply to $D_4$, and so we shall ignore the possibility of a hidden sector on $D_4$ since it would have the same features of the hidden sector on $D_3$. On the other hand, the SM degrees of freedom can be charged under the gauge group on $D_2$ since there is an intersection among $D_1$ and $D_2$. However this would still be a hidden sector since $\tau_2$ is a big cycle, and so the corresponding gauge coupling would be hyper-weak. Thus the relevant couplings are (see App. \ref{AppCanNorm} for the details of canonical normalisation): 
\be
\mc{L}_{\rm ax-gauge} = \frac{a}{M_p}\left[\mu_1\, \tilde F_{\rm vis} F_{\rm vis} + \mu_2 \frac{\langle\tau_3\rangle^{3/2}}{\vov}\, \tilde{F}_2 F_2 + \mu_3 \left(\frac{m_a}{m_{\theta_3}}\right)^2\tilde{F}_3 F_3 \right] + \mu_4 \frac{a_2}{M_p} \tilde{F}_2 F_2\,,
\label{fib_canon_couplings_aniso}
\ee
where the $\mu_i$'s are $\mc{O}(1)$ numerical coefficients. Notice that the QCD axion $a$ has a standard Planckian coupling to visible gauge bosons since it arises from a bulk cycle. On the other hand its coupling to the hidden gauge bosons on $D_2$ is $\vo$-suppressed, while $a$ is essentially decoupled from the hidden sector on $D_3$ since $(m_a/m_{\theta_3})^2\propto (\Lambda_\QCD/M_p)^4\simeq 10^{-76}$. The ALP $a_2$ features instead an $\mc{O}(1/M_p)$ coupling to hidden gauge bosons on $D_2$ but it is decoupled from the other sectors. These results are again due to the fact that $a_2$ is essentially massless and $D_3$ has no intersection with $D_1$ and $D_2$. 

\subsection{An example with arbitrary $h^{1,1}$}
\label{Largeh11}

A generic CY threefold is characterised by hundreds of K\"ahler moduli, i.e. $h^{1,1}\sim\mc{O}(100)$, and so one may wonder whether the axion physics of this more complicated case would display features similar to the ones of the relatively simple case with $h^{1,1}=4$ analysed above. As we have stresses, this depends on the details of moduli stabilisation. In this section we shall describe how to freeze all K\"ahler moduli for arbitrary $h^{1,1}$ following \cite{Cicoli:2016chb}. We will obtain an LVS vacuum where $\vo$ can be taken large enough to trust the $\alpha'$ expansion. 

The only requirement on the geometry is the presence of 2 blow-up modes, the first, $D_\SM$, to host the SM and the second, $D_{\rm np}$, to support non-perturbative effects. This condition is not too restrictive since del Pezzo divisors arise very frequently in CY constructions. We shall therefore consider an internal volume of the form: 
\be
\vo = \frac16 \sum_{i,j,k=1}^{N} k_{ijk} t_i t_j t_k - \gamma_\SM \tau_\SM^{3/2}- \gamma_{\rm np} \tau_{\rm np}^{3/2}\,,\qquad \text{with}\quad N=h^{1,1}-2\gg 1\,.
\label{Volnew}
\ee
As explained in Sec. \ref{ModStabLead}, the leading contributions to the scalar potential in a large-$\vo$ expansion arise from $\mc{O}(\alpha'^3)$ corrections to $K$ and $T_{\rm np}$-dependent non-perturbative corrections to $W$ as in (\ref{KandWcorr}), which stabilise $\tau_{\rm np}\sim g_s^{-1}$, $\theta_{\rm np}\sim \pi/a_{\rm np}$ and $\vo\sim e^{1/g_s}$. Similarly to the isotropic case studied in Sec. \ref{IsotropicLimit}, the SM cycle $\tau_\SM$ is instead fixed by the interplay of D-terms, F-terms of matter fields and $\tau_\SM$-dependent loop corrections. This stabilisation procedure ensures that the internal volume can be exponentially large while $\tau_\SM = \alpha_\SM^{-1}\sim \mc{O}(10-100)$ can reproduce the observed value of the SM gauge coupling. Moreover the axion $\theta_\SM$ behaves as a perfect QCD axion candidate with a decay constant of order the string scale.

At this level of approximation, there are still $(N-1)$ saxionic and $N$ axionic flat directions (without considering the QCD axion $\theta_\SM$ which we assume to be lifted by QCD instantons). All the $(N-1)$ flat saxions can be lifted by including subdominant $\alpha'$ effects. In 10D the first higher derivative corrections which modify the 4D scalar potential upon dimensional reduction, arise at $\mc{O}(\alpha'^3)$ and scale as $G_3^2 R^3$. In 4D they generate the term proportional to $\xi$ in (\ref{LVSpot}). Additional 10D $\mc{O}(\alpha'^3)$ terms scale as $G_3^4 R^2$, $G_3^6 R$ and $G_3^8$, and they give rise in 4D to higher F-term contributions to the scalar potential which scale respectively as $F^4$, $F^6$ and $F^8$ \cite{Ciupke:2015msa}. When the superspace derivative expansion is under control \cite{Cicoli:2013swa}, these terms represent just negligible corrections to the LVS potential (\ref{LVSpot}). However they can be the leading effects to lift any remaining flat direction. In particular, the form of $\mc{O}(\alpha'^3)$ $F^4$ corrections for an arbitrary CY $X$ has been determined to be \cite{Ciupke:2015msa} (ignoring the dependence on del Pezzo moduli):
\be
V_{F^4} = \frac{\lambda W_0^4}{\vo^4}\sum_{i=1}^N \Pi_i t_i\,,
\ee
where $\lambda \propto g_s^{-1/2}$ is a positive coefficient \cite{Grimm:2017okk} and the $\Pi_i$'s are $\mc{O}(1)$ topological quantities which can be expressed in terms of the second Chern class $c_2$ as $\Pi_i = \int_X c_2 \wedge \hat{D}_i$ \cite{Ciupke:2015msa}. Notice that $\Pi_i\geq 0$ $\forall i=1,...,N$ in a basis of the K\"ahler cone where $t_i \geq 0$. The total potential can thus be written schematically as:
\be
V_{\rm tot} = V_\LVS(\vo) + V_{F^4}(\vo, t_i)\,,
\ee
where we have highlighted the moduli-dependence of each contribution. Extremising with respect to the 2-cycle moduli, we obtain:
\be
\frac{\partial V_{\rm tot}}{\partial t_i} = \left(\frac{\partial V_\LVS}{\partial \vo} -\frac{4\lambda W_0^4 \Pi_k t_k}{\vo^5}\right)\tau_i+ \frac{\lambda W_0^4 \Pi_i}{\vo^4}\,.
\label{Vi}
\ee
Using $t_i\tau_i = 3\vo$, it is easy to realise that $t_i \partial_{t_i} V_{\rm tot}=0$ implies:
\be
\frac{\partial V_\LVS}{\partial \vo} = \frac{11\lambda W_0^4}{3 \vo^5}\,\Pi_i t_i\,.
\label{Vint}
\ee
Plugging this result in (\ref{Vi}) we find:
\be
\frac{\Pi_k t_k}{3\vo}=\frac{\Pi_i}{\tau_i}\qquad \forall i=1,...,N\,.
\ee
This relation fixes $(N-1)$ moduli in terms of one of them, say $\tau_N$, as:
\be
\tau_j = \frac{\Pi_j}{\Pi_N}\,\tau_N\,,\qquad \forall j =1,..., N-1\,.
\label{MinF4}
\ee
The positivity of $\lambda$ and the $\Pi_j$'s ensures that this is a well-behaved minimum \cite{Cicoli:2016chb}. Substituting (\ref{MinF4}) in (\ref{Volnew}), we obtain: 
\be
\tau_N = h_N(k_{ijk},\Pi_i)\,\vo^{2/3}\,,\qquad\Rightarrow\qquad
\tau_i = h_i(k_{ijk},\Pi_i)\,\vo^{2/3}\,,\qquad \forall i=1,..., N\,,
\label{tauN}
\ee
where $h_i(k_{ijk},\Pi_i)$ are functions of the intersection numbers and the topological quantities $\Pi_i$. The overall volume $\vo$ is fixed by solving (\ref{Vint}) which would yield just a subleading shift of the standard LVS solution (\ref{swiss_minimum}). For $\vo\sim e^{1/g_s}$ and $h_i\sim \mc{O}(1-10)$ $\forall i$, the minimum (\ref{tauN}) leads to an isotropic CY where all divisor volumes are large enough to trust the $\alpha'$ expansion.

Given that this stabilisation is purely perturbative, at this level of approximation $N$ axions are still flat. They can be lifted by including non-perturbative corrections to $W$ which however tend naturally to give rise to axion masses below the present Hubble constant for $\vo\gtrsim (h^{1,1})^7\gtrsim 10^{14}$. Let us stress that for such a large value of $\vo$ the SM is naturally expected to be supported on a blow-up mode since matching $\tau_* =\alpha_\SM^{-1}\sim\mc{O}(10$-$100)$ for a bulk cycle $\tau_*$ would need from (\ref{MinF4}) a very unnatural hierarchy between $\Pi_*$ and $\Pi_N$ of order $10^{-8}$ for $\tau_N\sim \vo^{2/3}\sim 10^{10}$.

\subsubsection*{Mass spectrum and decay constants}

The mass of the 3 moduli fixed at leading order, $\vo$, $\tau_{\rm np}$ and $\theta_{\rm np}$ is given in (\ref{LeadMass}). The mass spectrum of the remaining moduli becomes:
\bea
m_{\tau_\SM}&\simeq& m_{3/2}\,,\qquad
m_{\tau_j}\simeq m_{3/2}\left(\frac{m_{3/2}}{M_p}\right)^{5/6}\,, \qquad\forall j=1,...,N-1\,,
\nn \\
m_{\theta_\SM}&\equiv& m_a \simeq \frac{\Lambda_\QCD^2}{f_{\theta_\SM}}\,,
\qquad m_{\theta_i} \simeq M_p\,e^{-\pi\tau_i/\mathfrak{n}_i}\sim 0
\,,\qquad\forall i=1,...,N\,,
\label{m_isoNew}
\eea
where all the ALPs $\theta_i$'s are essentially massless. The decay constants scale as in the isotropic case with $h^{1,1}=4$ analysed in Sec. \ref{IsotropicLimit}:
\be
f_{\theta_\SM}\equiv f_a =\frac{c_\SM}{\langle\tau_\SM\rangle^{1/4}}\frac{M_p}{\sqrt{\langle\vo\rangle}}\,,\qquad f_{\theta_i}=c_i\frac{M_p}{\langle\tau_i\rangle} = \frac{c_i}{h_i}\frac{M_p}{\langle\vo\rangle^{2/3}}\,,\quad \forall i =1,...,N\,,
\label{f_isoNew}
\ee
where $c_\SM$ and the $c_i$'s are $\mc{O}(1)$ moduli-independent coefficients. The decay constant of the QCD axion $\theta_\SM$ scales again as the string scale, whereas the decay constant of each ultra-light ALP is controlled by the Kaluza-Klein scale. Contrary to the case with $h^{1,1}=4$ where values of $\vo$ of order $\vo\sim\mc{O}(10^3-10^4)$ could still be compatible with an EFT under control, for $h^{1,1}\sim\mc{O}(100)$ we should focus only on the region $\vo\gtrsim \mc{O}(10^{14})$. Thus we are naturally led to the region with TeV-scale supersymmetry and an intermediate scale QCD axion decay constant. 

\subsubsection*{Axion couplings}

We assume that the SM can be realised with a stack of magnetised D7-branes wrapped around $D_\SM$. On the other hand, the `big' divisors $D_i$, $i=1,...,N$, can in principle host several hidden sectors. The coupling of the QCD axion and the $N$ ultra-light ALPs to visible and hidden gauge bosons can be derived from the moduli-dependence of the corresponding gauge kinetic functions. Denoting the canonically normalised QCD axion as $a$, the ALPs as $a_i$ (the results of App. \ref{AppCanNorm} can be easily generalised to the isotropic case with many bulk K\"ahler moduli), and the field strengths as $F_{\rm vis}$ and $F_i$, we end up with: 
\be
\mc{L}_{\rm ax-gauge} =  \frac{a}{f_a}\left[\frac{\lambda_\SM}{\langle\tau_\SM\rangle}\, \tilde F_{\rm vis} F_{\rm vis} + \frac{\sqrt{\langle\tau_\SM\rangle}}{\vov} \sum_{i=1}^N\tilde\lambda_i\tilde{F}_i F_i  \right] 
+ \sum_{i=1}^N\hat\lambda_i\frac{a_i}{M_p} \tilde{F}_i F_i\,,
\label{IsoCouplings}
\ee
where again $\lambda_\SM$, the $\tilde\lambda_i$'s and the $\hat\lambda_i$'s are numerical $\mc{O}(1)$ coefficients. The coupling of the QCD axion $a$ to visible gauge bosons is enhanced with respect to $1/M_p$ while the coupling to hidden degrees of freedom is $\vo$-suppressed. This is due again to the fact that $D_\SM$ is a shrinkable del Pezzo divisor and the vanishing of the mass of the $N$ ALPs.

\section{Statistics of axion physics in the flux landscape}
\label{sec:Pheno}

Building on our previous results \cite{Broeckel:2020fdz}, we investigate the statistical distribution in the type IIB flux landscape of various quantities of axion physics which are phenomenologically interesting. To this end, we first express these quantities in terms of the microscopic parameters, as we did in Sec. \ref{sec:LVS} via moduli stabilisation, and we then exploit their distributions. In particular the shall consider the following distributions for the underlying flux-dependent parameters $g_s$ and $W_0$, and the rank $\mathfrak{n}$ of the condensing gauge group which generates non-perturbative corrections to the superpotential:
\bi
\item The distribution of the string coupling $g_s$ is taken to be uniform. This result was explicitly checked in \cite{Broeckel:2020fdz} for rigid CY manifolds and is believed to hold for more general cases as well \cite{Blanco-Pillado:2020wjn}. Hence in the following we shall take $dN \simeq  dg_s$ where $N$ is the number of flux vacua.

\item Based on the seminal work \cite{Denef:2004ze}, the tree-level superpotential $W_0$ is assumed to be uniformly distributed as a complex variable, resulting in $dN \simeq |W_0| d|W_0|$. Note that this distribution might be different in regions where $|W_0|$ is exponentially small since recent constructions of KKLT vacua obtained $|W_0|\sim e^{-1/g_s}$ \cite{Demirtas:2019sip, Demirtas:2020ffz, Blumenhagen:2020ire}. However, as explained in Sec. \ref{Intro}, KKLT vacua feature only heavy axions with a mass of order $m_{3/2}$, and so we shall focus just on regions where $|W_0|\sim \mc{O}(1$-$10$) where its distribution can be taken as uniform. 

\item The distribution of the rank of the condensing gauge group $\mathfrak{n}$ in the type IIB flux landscape is still poorly understood. All globally consistent type IIB CY models which have been constructed so far feature contributions to the superpotential which arise from just gaugino condensation in a pure $SO(8)$ sector (corresponding to $\mathfrak{n} = 6$) and ED3 instantons (with $\mathfrak{n} = 1$). It is therefore still unclear if an actual distribution of $\mathfrak{n}$ exists. If so, we argue that it should scale as $dN\simeq -\mathfrak{n}^{-r} d\mathfrak{n}$ with $r>0$, since the number of flux vacua $N$ is expected to decrease when $\mathfrak{n}$ increases as D7-tadpole cancellation is easier to satisfy for smaller values of $\mathfrak{n}$.  
\ei

\subsection{Axion decay constants}

Let us start with the axion decay constants. After evaluating the decay constants at the minimum of the scalar potential, we compute their distributions in the flux landscape using the scaling of the number of vacua $N$ with the underlying parameters $g_s$, $W_0$ and $\mathfrak{n}$.

\subsubsection*{Isotropic limit}

The axion decay constants in the isotropic limit are given in (\ref{f_iso}). Being exponentially large, the main quantity which controls their distribution is the overall volume $\vo$. Using (\ref{swiss_minimum}), we can therefore approximate the axion decay constants as:
\be
f_a \sim M_p\,e^{- \frac{c}{g_s\mathfrak{n}_4}}\,,\qquad 
f_{\theta_1} \sim f_{\theta_2} \sim M_p\,e^{- \frac{4c}{3g_s\mathfrak{n}_4}}\qquad\text{with}\quad c = \pi \left(\frac{\xi}{2\alpha} \right)^{2/3}\,.
\ee
Notice that, at leading order, the decay constants do not depend on $W_0$. Hence we can vary them with respect to just $g_s$ and $\mathfrak{n}_4$, obtaining:
\bea
d f &=& \frac{\partial f}{\partial g_s}\,d g_s + \frac{\partial f}{\partial \mathfrak{n}_4}\, d \mathfrak{n}_4
\simeq \frac{f}{(g_s \mathfrak{n}_4)^2} \left(\mathfrak{n}_4\,d g_s + g_s \,d \mathfrak{n}_4 \right) \nn \\
&\simeq& f \left[\ln\left(\frac{M_p}{f}\right)\right]^2 \left(\mathfrak{n}_4\,d g_s + g_s \,d \mathfrak{n}_4 \right),
\label{fadistr}
\eea
where $f$ can be any of the 3 decay constants, $f_a$, $f_{\theta_1}$ and $f_{\theta_2}$, and in the last step we have introduced Planck units. Using $d g_s \simeq d N$ and $d N \simeq -\mathfrak{n}_4^{-r} \,d \mathfrak{n}_4$ with $r>0$, (\ref{fadistr}) takes the form:
\be
d f \simeq \mathfrak{n}_4\,f \left[\ln\left(\frac{M_p}{f}\right)\right]^2  \left[1 - \frac{\tilde{c} \,\mathfrak{n}_4^{r-2}}{\ln\left(\frac{M_p}{f}\right)} \right] d N\,,
\label{fadistrib}
\ee
where $\tilde{c}=c$ for $f_a$ and $\tilde{c}=4 c/3$ for $f_{\theta_1}$ and $f_{\theta_2}$. Ignoring subdominant logarithmic effects, we therefore obtain that the distributions of the decay constants of both the QCD axion and the 2 ultra-light ALPs scale as:
\be
N (f_a) \sim  \ln\left(\frac{f_a}{M_p}\right)\qquad\text{and}\qquad
N (f_{\theta_i}) \sim  \ln\left(\frac{f_{\theta_i}}{M_p}\right),\quad i=1,2\,.
\label{faDistr}
\ee
Let us stress 3 important points:
\ben
\item Isotropic models with the SM localised on a blow-up cycle feature only a mild logarithmic preference for higher values of the axion decay constants.

\item As can be seen from (\ref{fadistrib}), for $f\ll M_p$ and $0< r\leq 2$, the distribution of $f$ is driven mainly by the distribution of $g_s$. Moreover the final result (\ref{faDistr}) is unchanged if the distribution of $g_s$ is taken to be power-law.

\item As can be seen again from (\ref{fadistrib}), the unknown distribution of $\mathfrak{n}_4$ would start being important only for $r>2$. However it is reassuring to notice that it would affect only the form of subleading logarithmic corrections to (\ref{faDistr}). 
\een

\subsubsection*{Anisotropic limit}

For the anisotropic case the axion decay constants are given in (\ref{f_aniso}). As already observed in Sec. \ref{AnisSM}, the QCD axion decay constant is fixed around the GUT scale, $f_a\sim M_\GUT$, by the need to match the correct SM gauge coupling. If this phenomenological condition is dropped, however $f_a$ would feature a logarithmic distribution as in (\ref{faDistr}). The same is true for the distribution of the decay constant of the ALP $\theta_2$. However in this case the phenomenological requirement $\alpha_\SM^{-1} = \tau_1 \simeq \mathcal{O}(10$-$100$) still leaves some freedom to vary $f_{\theta_2}$ in the flux landscape since from (\ref{f_aniso}) we have $f_{\theta_2} \simeq n_2 g_s^2 M_\GUT$. Notice that the ALP decay constant does not depend on $W_0$ which has however to respect the relation (\ref{W0gsRel}) to keep $\tau_1=\alpha_\SM^{-1}$ constant when $g_s$ is varied. Differentiating $f_{\theta_2}$ with respect to $g_s$ and $\mathfrak{n}_2$ we thus obtain:
\be
\frac{d f_{\theta_2}}{f_{\theta_2}} = \frac{\partial f_{\theta_2}}{\partial g_s}\,\frac{d g_s}{f_{\theta_2}} + \frac{\partial f_{\theta_2}}{\partial \mathfrak{n}_2}\, \frac{d \mathfrak{n}_2}{f_{\theta_2}}
= 2\,\frac{d g_s}{g_s} +\frac{d \mathfrak{n}_2}{\mathfrak{n}_2}\,.
\label{fa2distr}
\ee
Using again $d g_s \simeq d N$ and $d N \simeq -\mathfrak{n}_2^{-r} \,d \mathfrak{n}_2$ with $r>0$, (\ref{fa2distr}) reduces to:
\be
d f_{\theta_2} \simeq \sqrt{\mathfrak{n}_2}\,\sqrt{f_{\theta_2}\,M_\GUT} \left(1 - \frac{g_s}{2} \,\mathfrak{n}_2^{r-1}\right) d N\,.
\label{fa2distrib}
\ee
For $0<r\leq 1$, $\mathfrak{n}_2\geq 1$ and $g_s\ll 1$, the second term in brackets in (\ref{fa2distrib}) is always smaller than unity. This term could instead become larger than $1$ for $r>1$. However, for $g_s\lesssim 0.1$, this would require large values of 
$\mathfrak{n}_2$ which are hard to realise in explicit examples (the largest value obtained so far is $\mathfrak{n}_2=6$ for gaugino condensation in a pure $SO(8)$ gauge theory which would however still yield a second term of $\mc{O}(1)$ for $r\leq 3$). We shall therefore consider the term in brackets in (\ref{fa2distrib}) of order unity, and obtain:
\be
N (f_{\theta_2}) \sim  \sqrt{\frac{f_{\theta_2}}{M_\GUT}}\,.
\label{fa2Distr}
\ee
Let us make 2 important observations:
\ben
\item Interestingly, we obtained now a power-law distribution for the ALP decay constant whose scaling is however very similar to the logarithmic case due to the mild square root dependence.

\item The distribution (\ref{fa2Distr}) holds as long as $W_0$ can be tuned to satisfy the relation (\ref{W0gsRel}) which implies $W_0 \sim e^{-1/g_s}$. In the absence of a dynamical mechanism which fixes the flux superpotential in terms of dilaton-dependendent non-perturbative effects, this relation would however not hold anymore when $g_s$ is taken very small. A good estimate for the lowest value of the ALP decay constant for which (\ref{fa2Distr}) still applies, can be obtained for $g_s\simeq 0.01$ which would give $f_{\theta_2}\gtrsim 10^{12}$ GeV.
\een

\subsubsection*{Model with arbitrary $h^{1,1}$}

As shown in Sec. \ref{Largeh11}, the results of the isotropic case with the SM on a blow-up cycle can be generalised to models with an arbitrarily large number of K\"ahler moduli where all saxions can be explicitly stabilised by $\alpha'$ corrections to the scalar potential. In this case the axion decay constants are given in (\ref{f_isoNew}), and they scale with the CY volume as in the isotropic case discussed above. Hence we expect again a logarithmic distribution in the type IIB flux landscape as in (\ref{faDistr}):
\be
N (f_a) \sim  \ln\left(\frac{f_a}{M_p}\right)\qquad\text{and}\qquad
N (f_{\theta_i}) \sim  \ln\left(\frac{f_{\theta_i}}{M_p}\right),\quad \forall i=1,...,N\,.
\label{faDistrLargeh11}
\ee
Let us comment on the regime of validity of these distributions. They hold at fixed $h^{1,1}$ when moving in the K\"ahler moduli space by varying microscopic parameters like $g_s$ after the decay constants are written in terms of them thanks to moduli stabilisation. These results are complementary to the ones of \cite{Mehta:2020kwu,Mehta:2021pwf} (see also \cite{Halverson:2019cmy} for qualitatively similar findings) which found an approximate log-normal distribution for the axion decay constants of a given CY model focusing at the tip of the stretched K\"ahler cone. Moreover they found that the mean value of the $f_{\theta_i}$'s decreases when $h^{1,1}$ increases. 

Given that the tip of the stretched K\"ahler cone corresponds to the smallest values of the K\"ahler moduli which are compatible with a controlled $\alpha'$ expansion, and the axion decay constants are inversely proportional to 4-cycle volumes, the values of the $f_{\theta_i}$'s obtained by \cite{Mehta:2020kwu,Mehta:2021pwf} represent the largest values of the axion decay constants compatible with a trustable EFT. These values would therefore provide an upper bound for the regime of validity of our logarithmic distributions (\ref{faDistrLargeh11}) which can be integrated with the results of \cite{Mehta:2020kwu,Mehta:2021pwf} to describe how the number of flux vacua changes has a function of both $f_{\theta_i}$ and $h^{1,1}$. 

\begin{figure}[htb]
\begin{center}
\includegraphics[height=2.5in,width=2.95in]{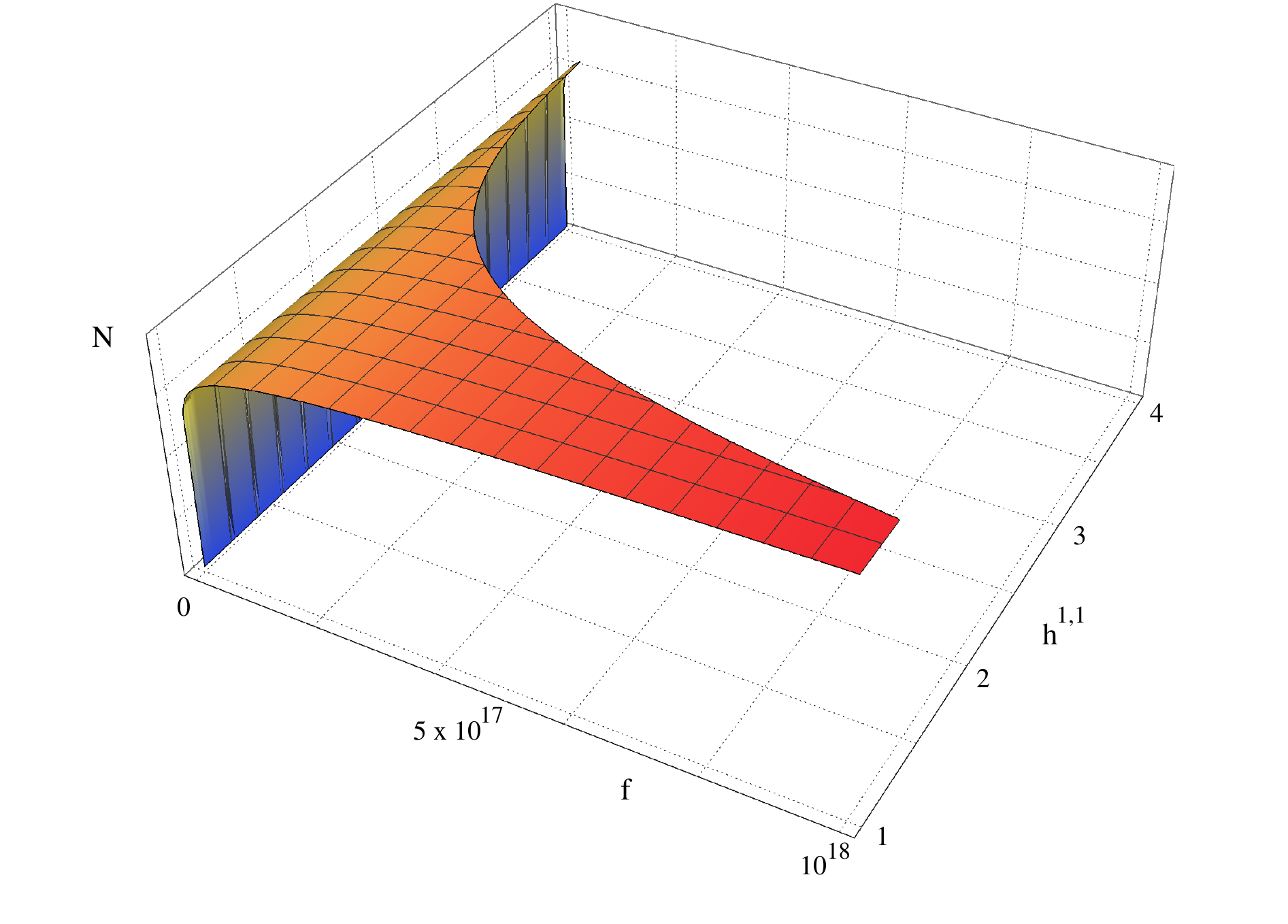}
\includegraphics[height=2.5in,width=2.95in]{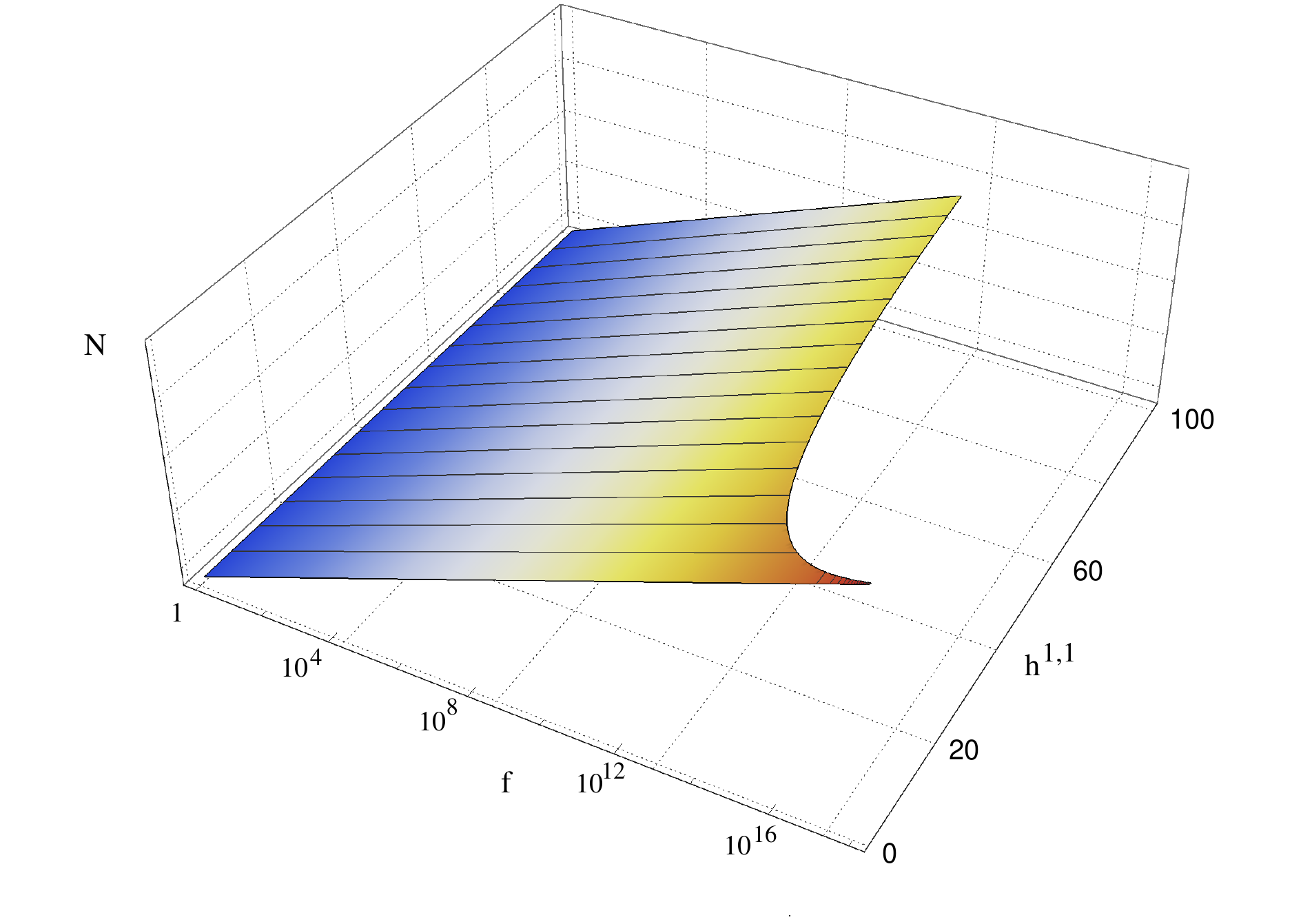}
\caption{Distribution of the number of flux vacua as a function of the mean value of the axion decay constants $f$ and $h^{1,1}$ with the constraint $f \lesssim M_p/(h^{1,1})^3$. At fixed $h^{1,1}$, $N(f)\simeq \ln(f/M_p)$. The plot on the right hand side shows a logarithmic scale for $f$.}
\label{Fig1}
\end{center}
\end{figure}

As an illustrative example, we consider the distribution $N(f,h^{1,1})$ where $f$ is the mean value of the axion decay constants. As derived in \cite{Demirtas:2018akl}, the requirement to trust the $\alpha'$ expansion implies that the volume of each 4-cycle grows with $h^{1,1}$ as $\tau_i\gtrsim (h^{1,1})^3$ $\forall i=1,...,h^{1,1}$ (at least for basis elements obtained from generators of the cone of effective divisors). On the other side, as we have seen in Sec. \ref{sec:LVS}, the axion decay constants scale as $f_{\theta_i}\simeq M_p/\tau_i$. Combining the two results gives a qualitative understanding of the fact the mean value $f$ of the log-normal distribution found in \cite{Mehta:2020kwu,Mehta:2021pwf} decreases as $h^{1,1}$ increases. Moreover, we can obtain an explicit estimate of the upper bound for our logarithmic distributions:
\be
f\lesssim f_{\rm max}(h^{1,1})\simeq \frac{M_p}{ (h^{1,1})^3}\,,
\ee
where, for a given $h^{1,1}$, $f$ can take different values by moving in the stretched K\"ahler cone in a way compatible with moduli stabilisation, and $f=f_{\rm max}$ at the tip. Hence we expect the following distribution for the number of type IIB flux vacua as a function of $f$ and $h^{1,1}$ (see Fig. \ref{Fig1}):
\be
N (f,h^{1,1}) \sim  \ln\left(\frac{f}{M_p}\right)\qquad\text{with}\qquad f\lesssim \frac{M_p}{ (h^{1,1})^3}\,.
\ee

\subsection{Axion masses}

Let us now compute the distribution of axion masses in the flux landscape. As in the previous section we first compute the differential of the masses and then use the known scaling of the parameters $g_s$, $W_0$ and $\mathfrak{n}$ in order to determine the distribution.

\subsubsection*{Isotropic limit}

The mass spectrum of the isotropic case with SM on a blow-up cycle is summarised in (\ref{m_iso}). Using (\ref{fadistrib}) which can be approximated as $df_a\simeq f_a\, dN$, the distribution of the mass of the QCD axion takes the form:
\be
d m_a \simeq - \frac{m_a}{f_a}\,d f_a \simeq - m_a \,dN\qquad\Rightarrow\qquad N(m_a) \sim -\ln\left(\frac{m_a}{M_p}\right).
\label{Nma}
\ee
On the other hand, the distributions of the masses of the two ultra-light ALPs can be easily derived by noticing that their masses can be expressed in terms of the corresponding decay constants as:
\be
m_{\theta_1}\simeq M_p\,e^{-\frac{M_p}{2\sqrt{2}f_{\theta_1}}}\qquad\text{and}\qquad 
m_{\theta_2}\simeq M_p\,e^{-\frac{M_p}{2f_{\theta_2}}}\,.
\label{mifi}
\ee
This result implies:
\be
d m_{\theta_i} \simeq m_{\theta_i} \frac{M_p}{f_{\theta_i}} \frac{d f_{\theta_i}}{f_{\theta_i}}\simeq m_{\theta_i} \ln\left(\frac{M_p}{m_{\theta_i}}\right) dN\qquad \forall\,i=1,2\,,
\label{dmthetai}
\ee
which yields the following distribution (neglecting subdominant logarithmic corrections):
\be
N (m_{\theta_i}) \sim \ln \left(\frac{m_{\theta_i}}{M_p} \right)\qquad \forall\,i=1,2\,.
\label{m12_distr_iso}
\ee
Notice that we find again a logarithmic distribution for the mass of both the QCD axion and the 2 ultra-light ALPs. However (\ref{Nma}) and (\ref{m12_distr_iso}) have a different sign, implying that, in the QCD axion case, the type IIB flux landscape has a mild logarithmic preference for low scale masses, while the ALP case features more vacua at large mass values.

\subsubsection*{Anisotropic limit}

The masses for the anisotropic geometry with SM on the bulk divisor $D_1$ are summarised in (\ref{mass_aniso}). As already pointed out in the previous section, this model is strongly constrained by the requirement to match the observed SM coupling. This sets the QCD axion decay constant around the GUT scale, $f_a\simeq M_\GUT$, without a distribution. Thus the QCD axion mass would also be fixed at $m_a \simeq \Lambda_\QCD^2/f_a \simeq 1$ neV. 

The mass of the ALP $\theta_2$ can instead take different values in the flux landscape with a distribution which is again logarithmic. This result can be easily inferred by first writing $m_{\theta_2}$ as in (\ref{mifi}) and then differentiating as below:
\be
d m_{\theta_2} \simeq m_{\theta_2} \frac{M_p}{f_{\theta_2}} \frac{d f_{\theta_2}}{f_{\theta_2}}\simeq m_{\theta_2} \left[\ln\left(\frac{M_p}{m_{\theta_2}}\right)\right]^{3/2} dN\,,
\label{dm2}
\ee
where we used (\ref{fa2distr}) approximated as:
\be
\frac{d f_{\theta_2}}{f_{\theta_2}} \simeq \frac{d g_s}{g_s}\simeq \sqrt{\frac{M_p}{f_{\theta_2}}}\,dN
\simeq \sqrt{\ln\left(\frac{M_p}{m_{\theta_2}}\right)}\,dN\,.
\ee
Barring subleading logarithmic effects, (\ref{dm2}) therefore implies: 
\be
N (m_{\theta_2}) \sim \ln \left(\frac{m_{\theta_2}}{M_p} \right).
\ee
This distribution, similarly to the one of $f_{\theta_2}$ derived in (\ref{fa2Distr}), holds as long as $W_0$ can be tuned to satisfy the condition (\ref{W0gsRel}) which keeps $\tau_1 =\alpha_\SM^{-1}$ fixed at the right SM gauge coupling. In the case of the ALP decay constant, we estimated that its distribution is valid for $f_{\theta_2}\gtrsim 10^{12}$ GeV. Using (\ref{mifi}), this gives however a lower bound for the ALP mass which can be safely ignored since it would be much smaller than today's Hubble constant: $m_{\theta_2}\gtrsim e^{-10^6}\, M_p \ll H_0$.

\subsubsection*{Model with arbitrary $h^{1,1}$}

The mass spectrum for the model with a generic number of K\"ahler moduli is given in (\ref{m_isoNew}). Following the discussion of the distribution of the axion decay constants, the results for the distribution of the axion masses for the model with arbitrary $h^{1,1}$ would again be qualitatively similar to the isotropic case. Hence we expect logarithmic distributions of the form:
\be
N(m_a) \sim -\ln\left(\frac{m_a}{M_p}\right)
\qquad\text{and}\qquad
N (m_{\theta_i}) \sim \ln \left(\frac{m_{\theta_i}}{M_p} \right)\qquad \forall\,i=1,...,N\,.
\label{mi_distr}
\ee
As for the case of the axion decay constants discussed above, the results of \cite{Mehta:2020kwu,Mehta:2021pwf} can be combined with ours to give an upper bound for the regime of validity of the distributions of the ALP masses as a function of $h^{1,1}$, i.e. $m_{\theta_i}\lesssim m_{\theta_i}^{\rm max}(h^{1,1})$.

\subsection{Dark matter abundance}

Let us now study the distribution of the axion DM abundance produced via the standard misalignment mechanism. We distinguish between the case where the DM particle is the QCD axion and the case where it is an ultra-light ALP. In the QCD axion case, the DM abundance is given by:
\be
\frac{\Omega_\QCD h^2}{0.112} \simeq 6.3 \cdot \left(\frac{f_a}{10^{12}\,{\rm GeV}} \right)^{7/6} \left(\frac{\theta_{\rm in}}{\pi} \right)^2,
\label{QCDCDM}
\ee
while for the case of an ALP $\theta_i$, it reads:
\be
\frac{\Omega_\ALP h^2}{0.112} \simeq 1.4 \cdot \left(\frac{m_{\theta_i}}{1\,{\rm eV}} \right)^{1/2}\left(\frac{f_{\theta_i}}{10^{11}\,{\rm GeV}} \right)^2 \left(\frac{\theta_{i,{\rm in}}}{\pi} \right)^2.
\label{nonQCDCDM}
\ee
For natural $\mc{O}(\pi)$ values of the initial misalignment angles $\theta_{\rm in}$ and $\theta_{i,{\rm in}}$, the QCD axion can reproduce the observed DM adundance for $f_a\simeq 10^{11}$ GeV, while an ALP would require $m_{\theta_i} \simeq 5\cdot 10^{-21}$ eV for $f_{\theta_i} \simeq 10^{16}$ GeV (see App. \ref{abun} for a detailed scan through the underlying parameter space and some benchmark points).

\subsubsection*{Isotropic limit and model with arbitrary $h^{1,1}$}

In the isotropic case with SM on a blow-up cycle, the distribution of QCD axion DM abundance can be computed deriving (\ref{QCDCDM}) with respect to $f_a$ and then using the result (\ref{fadistrib}) which, at first approximation, implies $d f_a \simeq f_a\, dN$. Hence we end up with:
\be
d \left(\Omega_\QCD h^2\right) = \frac76 \left(\Omega_\QCD h^2\right)\frac{d f_a}{f_a} \simeq \left(\Omega_\QCD h^2\right) dN\,,
\ee
which gives:
\be
N \left(\Omega_\QCD h^2\right)\sim \ln \left(\Omega_\QCD h^2\right)
\ee
This result is very important since it implies that the number of type IIB flux vacua which can reproduce the correct value of the DM abundance for $\theta_{\rm in}\sim \mc{O}(\pi)$ is only logarithmically suppressed with respect to what has been considered so far as the typical stringy case with $f_a\sim M_\GUT$ and a tuned initial misalignment angle. 

The isotropic case features 2 ultra-light ALPs, $\theta_1$ and $\theta_2$. Both of them can behave as cold DM. Noticing from (\ref{mifi}) that the microscopic model sets a correlation between $m_{\theta_i}$ and $f_{\theta_i}$ of the form $m_{\theta_i}\simeq M_p\,e^{-\beta_i \frac{M_p}{f_{\theta_i}}}$ with $\beta_i\sim\mc{O}(1)$ $\forall i=1,2$, the distribution of the ALP DM abundance (\ref{nonQCDCDM}) is mainly controlled by $m_{\theta_i}$. We can therefore derive (\ref{nonQCDCDM}) just with respect to $m_{\theta_i}$ and obtain:
\be
d \left(\Omega_\ALP h^2\right) = \left[\frac12 \frac{d m_{\theta_i}}{m_{\theta_i}}+2 \frac{d f_{\theta_i}}{f_{\theta_i}}\right]\left(\Omega_\ALP h^2\right)\simeq  \frac{d m_{\theta_i}}{m_{\theta_i}} \left(\Omega_\ALP h^2\right) \simeq \left(\Omega_\ALP h^2\right) dN\quad\forall\,i=1,2\,,
\ee
where we used (\ref{dmthetai}) approximated as $d m_{\theta_i}\simeq m_{\theta_i}\,dN$. This implies for both $\theta_1$ and $\theta_2$:
\be
N \left(\Omega_\ALP h^2\right)\sim \ln \left(\Omega_\ALP h^2\right).
\ee
Thus we realise that also the distribution of the ALP DM abundance features a logarithmic behaviour. As already explained, this result should apply also to the distribution of the DM abundance of each ultra-light ALP of the model with arbitrarily large $h^{1,1}$.

\subsubsection*{Anisotropic limit}

In the anisotropic limit with the SM on the fibre divisor, the value of the QCD axion decay constant is fixed around the GUT scale once we focus just on vacua which match the SM coupling. Hence this would represent a typical stringy case which has been considered as `anthropic' since for $f_a \sim M_\GUT\sim 10^{16}$ GeV (\ref{QCDCDM}) would reproduce the correct DM abundance only for $\theta_{\rm in}\sim 0.001 \pi$.

The DM abundance associated to the ultra-light ALP $\theta_2$ would instead be distributed as:
\be
d \left(\Omega_\ALP h^2\right) \simeq  \frac{d m_{\theta_2}}{m_{\theta_2}}\left(\Omega_\ALP h^2\right) \simeq \left(\Omega_\ALP h^2\right) dN\quad\Rightarrow\quad N(\Omega_\ALP h^2) \sim \ln\left(\Omega_\ALP h^2 \right),
\ee
where we used $d m_{\theta_2}\simeq m_{\theta_2}\,dN$ from (\ref{dm2}). Similarly to the isotropic case, we find again a logarithmic distribution with however the difference, as already pointed out, that in the anisotropic case all expressions, (\ref{nonQCDCDM}) included,  are valid only for $f_{\theta_2}\gtrsim 10^{12}$ GeV (while we have seen that any value of $m_{\theta_2}$ is allowed).

\subsection{Axion couplings to gauge bosons}

Let us now study the distribution of the couplings between axions and gauge fields. Following the analysis of the previous sections, we first evaluate the couplings at the minimum of the scalar potential and we then determine their distribution in the flux landscape.

\subsubsection*{Isotropic limit and model with arbitrary $h^{1,1}$}

We start with the couplings in the isotropic case which are summarised in (\ref{fib_canon_couplings_iso}). The axion couplings to visible and hidden gauge bosons feature a similar behaviour also in the model with arbitrarily many K\"ahler moduli, as can be seen from (\ref{IsoCouplings}). Hence the results which we will obtain for the isotropic case can be directly extended to the more general case with arbitrarily large $h^{1,1}$ and SM built with a stack of D7-branes wrapped around a local del Pezzo divisor.

Interestingly, each ultra-light ALP couples in practice just to the corresponding hidden gauge fields with a coupling that is fixed at Planckian strength without a distribution in the landscape. This is a typical stringy behaviour, as expected for the imaginary part of a standard bulk modulus. On the other hand, the coupling between the QCD axion $a$ and SM gauge fields $\gamma$ is controlled by the string scale $M_s \sim M_p/\sqrt{\vo}$ since it is inversely proportional to $f_a$:
\be
g_{a \gamma \gamma} = \frac{\lambda_1}{\langle \tau_3 \rangle \,f_a}\sim \frac{1}{M_s}\,.
\ee
Thus the distribution of $g_{a\gamma\gamma}$ takes the form:
\be
d g_{a\gamma\gamma} \simeq - g_{a\gamma\gamma}\,\frac{d f_a}{f_a} \simeq - g_{a\gamma\gamma} \,dN\qquad\Rightarrow\qquad N(g_{a\gamma\gamma}) \sim -\ln\left(g_{a\gamma\gamma}\right).
\label{Nga}
\ee
where we used (\ref{fadistrib}) approximated as $df_a\simeq f_a\, dN$. Notice the mild logarithmic preference for smaller couplings. The coupling of the QCD axion to hidden gauge bosons $\gamma_{\rm h}$ living on stacks of D7-branes wrapped around the bulk divisors $D_1$ and $D_2$, is instead weaker than Planckian since these two divisors do not intersect with the del Pezzo 4-cycle $D_3$:
\be
g_{a \gamma_{\rm h}\gamma_{\rm h}} = \frac{\lambda_i
\sqrt{\langle \tau_3 \rangle}}{ \vov\, f_a} \sim \left(\frac{f_a}{M_p}\right) \frac{1}{M_p}\ll \frac{1}{M_p}\qquad \forall i=1,2\,.
\ee
Hence, using $df_a\simeq f_a\, dN \simeq g_{a\gamma_{\rm h}\gamma_{\rm h}}\, dN$, the distribution of the coupling between the QCD axion and hidden gauge bosons scales as:
\be
d g_{a\gamma_{\rm h}\gamma_{\rm h}} \simeq d f_a  \simeq g_{a\gamma_{\rm h}\gamma_{\rm h}} \,dN\qquad\Rightarrow\qquad N(g_{a\gamma_{\rm h}\gamma_{\rm h}}) \sim \ln\left(g_{a\gamma_{\rm h}\gamma_{\rm h}}\right).
\label{Ngah}
\ee
Contrary to the coupling to visible gauge fields, in this case the flux landscape features a logarithmic preference for larger couplings.

\subsubsection*{Anisotropic limit}

The axion-gauge couplings for the anisotropic case are summarised in (\ref{fib_canon_couplings_aniso}). Contrary to the isotropic case, the coupling of the QCD axion to visible sector gauge fields does not show a distribution since it is fixed at $1/M_p$, as typical of a string modulus. The difference with the isotropic case in this regard is due to the different topological origin of the QCD axion which in the isotropic case arises from the reduction of $C_4$ on a local del Pezzo 4-cycle while in the anisotropic case it is associated to the bulk divisor $D_1$. The behaviour of the ultra-light ALP $a_2$ is instead similar to the one of the 2 ALPs in the isotropic case since $a_2$ couples just to hidden degrees of freedom on $D_2$ with a fixed strength of order $1/M_p$. 

In the anisotropic case, the only couplings which can take different values in the flux landscape are the couplings of the QCD axion to the gauge bosons of the hidden sectors on $D_2$ and $D_3$, which we denote respectively as $\gamma_2$ and $\gamma_3$. The coupling $g_{a \gamma_2 \gamma_2}$ scales as:
\be
g_{a \gamma_2 \gamma_2} = \frac{\mu_2}{M_p} \frac{\langle\tau_3\rangle^{3/2}}{\vov} \sim \left(\frac{f_{\theta_2}}{M_p}\right) \frac{1}{M_p}\,.
\ee
As we have already estimated, in this case $10^{12}\,{\rm GeV}\lesssim f_{\theta_2}\lesssim 10^{16}$ GeV, which implies $10^{-6}\lesssim g_{a \gamma_2 \gamma_2}\,M_p\lesssim 10^{-2}$. In this regime of validity, the distribution of the  coupling $g_{a \gamma_2 \gamma_2}$ turns out to be:
\be
d g_{a \gamma_2 \gamma_2} \simeq  d f_{\theta_2}\simeq 
\sqrt{g_{a \gamma_2 \gamma_2}}\,  d N\qquad\Rightarrow\qquad
N (g_{a \gamma_2 \gamma_2}) \sim  \sqrt{g_{a \gamma_2 \gamma_2}}\,,
\label{ga2Distr}
\ee
where $d f_{\theta_2} \simeq \sqrt{f_{\theta_2}}\, d N$ from (\ref{fa2distrib}). Hence the coupling of the QCD axion to hidden gauge fields on $D_2$ is weaker than Planckian with a mild (due to the square root) power-law preference for couplings close to $0.01/M_p$. The QCD axion is instead almost decoupled from the degrees of freedom of the hidden D7-stack wrapping $D_3$ since $g_{a \gamma_3 \gamma_3}$ scales as:
\be
g_{a \gamma_3 \gamma_3} = \frac{\mu_3}{M_p} \left(\frac{m_a}{m_{\theta_3}}\right)^2 \sim \left( \frac{\Lambda_\QCD^2}{M_\GUT f_{\theta_2}}\right)^2\frac{\alpha_\SM^{-1}}{M_p}\,,
\ee
where $m_{\theta_3}\simeq m_{3/2} \simeq \sqrt{\alpha_\SM}\,f_{\theta_2}$ from (\ref{mass_aniso}) and (\ref{6Dm32}), and $m_a\simeq \Lambda_\QCD^2/M_\GUT$. For $\alpha_\SM^{-1}\simeq 100$ and $10^{12}\,{\rm GeV}\lesssim f_{\theta_2}\lesssim 10^{16}$ GeV, this coupling would be of order $10^{-66}\lesssim g_{a \gamma_2 \gamma_2}\,M_p\lesssim 10^{-58}$, and so we can safely set it to zero in the whole flux landscape.

\subsection{Dark radiation in Fibre Inflation}

A generic feature of models where reheating occurs due to the decay of a closed string modulus is the production of ultra-light bulk axions which yield extra dark radiation \cite{Cicoli:2012aq, Higaki:2012ar,Hebecker:2014gka, Cicoli:2015bpq}. This happens also in the interesting case of type IIB Fibre Inflation models \cite{Cicoli:2008gp, Cicoli:2016xae,Cicoli:2017axo,Cicoli:2018tcq,Cicoli:2016chb, Cicoli:2020bao, Burgess:2016owb, Cicoli:2011ct} where the CY volume takes the same form as in (\ref{voform}) and the fibre modulus $\tau_1$ plays the role of the inflaton. The inflationary potential is generated by perturbative corrections to the K\"ahler potential and the CY volume is fixed around $\vo\simeq 10^3$-$10^4$ by the need to reproduce the observed amplitude of the density perturbations generated by inflaton fluctuations during inflation. In order to have an efficient production of SM degrees of freedom at reheating, the SM D7-stack has to wrap the fibre divisor $D_1$. Hence a viable realisation of Fibre Inflation models requires to focus on the anisotropic case.

The inflaton $\tau_1$ is the lightest K\"ahler modulus and its perturbative decay after the end of inflation produces SM particles together with the QCD axion $\theta_1$ and the ultra-light ALP $\theta_2$ which are both relativistic and yield a $g_s$-dependent contribution to the effective number of relativistic species $N_{\rm eff}$ \cite{Cicoli:2018cgu}. One can thus exploit the known distribution of $g_s$ to derive the distribution of extra dark radiation in the flux landscape of Fibre Inflation models. The amount of extra dark radiation is parameterised by $\Delta N_{\rm eff}$ which is determined by the ratio of the inflaton branching ratio into hidden and visible degrees of freedom \cite{Cicoli:2018cgu}:
\be
\Delta N_{\rm eff}=\frac{43}{7}\frac{\Gamma_{\rm hid}}{\Gamma_{\rm vis}}\left(\frac{g_*(T_{\rm dec})}{g_*(T_{\rm rh})} \right)^{1/3}\simeq \frac{0.6}{\gamma^2}\,,
\label{DNeff}
\ee
where the parameter $\gamma$ controls the coupling of the inflaton to visible sector gauge bosons and depends on the string coupling:
\be
\gamma= \alpha_\SM\tau_1=g_s^{4/3} \alpha_\SM \vo^{2/3}\,,
\label{gamma}
\ee
where we have used (\ref{tau1minNew}) with $\alpha=1$ in the the volume form (\ref{voform}). Let us stress that in Fibre Inflation models the CY volume is fixed around $\vo\simeq 10^3$ by the need to reproduce the observed amplitude of the density perturbations generated by inflaton fluctuations during inflation. Hence in (\ref{gamma}) $\vo$ should be considered as constant. When varying $g_s$, this can be achieved by an appropriate choice of $W_0$ (see (\ref{swiss_minimum})). Moreover $g_s$ should be varied by keeping the SM coupling fixed at its phenomenological value. Given that $\alpha_\SM$ reads:
\be
\alpha_\SM^{-1}=\tau_1- \frac{h(\mc{F})}{g_s} = \gamma \,\alpha_\SM^{-1} - \frac{h(\mc{F})}{g_s}\,,
\label{Rel}
\ee
where $h(\mc{F})\geq 0$ is a non-negative function of the intersection numbers and the gauge flux $\mc{F}$ on the SM D7-brane stack, this implies that any variation of $\gamma$ (by varying $g_s$) should be compensated by a suitable change of $h(\mc{F})$ by considering a different choice of $\mc{F}$ (if this is allowed by the discreteness of the gauge flux quanta and by tadpole cancellation). Notice that $h(\mc{F})$ vanishes for $\mc{F}=0$, implying from (\ref{Rel}) $\gamma=1$ and $\Delta N_{\rm eff}$ fixed at $\Delta N_{\rm eff}\simeq 0.6$ \cite{Cicoli:2018cgu}. However $h(\mc{F})>0$ for $\mc{F}\neq 0$, and so in this case $\Delta N_{\rm eff}$ features a distribution in the flux landscape due to its dependence on $g_s$. We can estimate the regime of validity of this distribution by setting in (\ref{gamma}) $\alpha_\SM^{-1}=25$, $\vo=5\cdot 10^3$ and $g_s\lesssim 0.25$ to trust perturbation theory, which gives $\Delta N_{\rm eff}\gtrsim 0.17$. We can also obtain an upper bound on $\Delta N_{\rm eff}$ by requiring $\tau_1\geq \alpha_\SM^{-1}$ from (\ref{Rel}) since $h(\mc{F})>0$. This gives $\gamma\geq 1$ from (\ref{gamma}), and so $\Delta N_{\rm eff}\lesssim 0.6$.

Varying now (\ref{DNeff}) with respect to $\gamma$ using (\ref{gamma}) and $dg_s\simeq dN$, we obtain:
\be
\frac{d(\Delta N_{\rm eff})}{\Delta N_{\rm eff}}\simeq - \frac{d\gamma}{\gamma} \simeq-\frac{d g_s}{g_s}\simeq -\Delta N_{\rm eff}^{3/8}\,dN
\qquad\Rightarrow\qquad N (\Delta N_{\rm eff})\sim \Delta N_{\rm eff}^{-3/8}\,.
\ee
which gives a power-law distribution for extra dark radiation:
\be
N (\Delta N_{\rm eff})\sim \Delta N_{\rm eff}^{-3/8}\qquad\text{for}\qquad 0.17\lesssim \Delta N_{\rm eff}\lesssim 0.6\,.
\ee
Interestingly we find that the flux landscape of Fibre Inflation models features more vacua around $\Delta N_{\rm eff}\simeq 0.17$ which helps to satisfy current bounds on extra relativistic species. We stress again that this distribution is valid only for values of $\Delta N_{\rm eff}$ corresponding to values of $g_s$ which are compatible with a choice of $h(\mc{F})$ that keeps $\alpha_\SM$ constant.

\section{Discussion and conclusions}
\label{Conclusions}

In this paper we studied the statistics of axion physics in the type IIB flux landscape focusing on the model-independent case of closed string axions coming from the dimensional reduction of $C_4$. We argued that a proper understanding of moduli stabilisation is crucial in order to derive the main features of the low-energy phenomenology of stringy axions. 

In KKLT-like scenarios all axions are as heavy as the corresponding saxions due to non-perturbative stabilisation. If the saxion masses are larger than $\mc{O}(50)$ TeV in order to avoid cosmological problems, each axion is thus too heavy to behave as the QCD axion or as a very light ALP for fuzzy DM. On the contrary, moduli stabilisation schemes which rely on perturbative corrections are characterised by axion masses which are exponentially suppressed with respect to saxion masses. This singles out LVS models as the best case scenarios for analysing axion physics since they also yield an exponentially large CY volume which allows to keep the EFT under control even for a large number of K\"ahler moduli.

Hence we focused on an LVS model with $h^{1,1}=4$ which is simple enough to perform moduli stabilisation in full detail but, at the same time, rich enough to show all the main features of axion physics which we consider to be valid in general for models with more K\"ahler moduli. We considered two regimes: ($i$) the isotropic limit with the SM on D7-branes wrapping a local blow-up cycle, and ($ii$) the anisotropic limit where the SM lives on a D7-stack wrapped around a bulk divisor. In both cases all phenomenologically interesting quantities, like axion decay constants, axion masses, contributions to the DM abundance and axion-gauge bosons couplings, feature a logarithmic distribution in the flux landscape. In the isotropic case however, the request to reproduce the correct SM gauge coupling selects a subset of the underlying parameter space where some distributions turn into a mild power-law behaviour.

Regarding the QCD axion, in the isotropic case it comes from the reduction of $C_4$ on a blow-up mode, whereas in the anisotropic case it is associated to a bulk cycle. In the last case its decay constant is fixed around the GUT scale by the need to match $\alpha_\SM$. On the other hand, in the first case $f_a$ is distributed logarithmically with just a mild preference for GUT-scale values in comparison with cases where $f_a$ is around intermediate scales. We consider this case to be more generic in the string landscape since realisations of the QCD axion from bulk cycles require an anisotropic moduli fixing which would require a good amount of tuning for relatively large values of $\vo$, while the case of a blow-up QCD axion can work with either isotropic or anisotropic models. We therefore conclude that what has been so far claimed to be the typical stringy situation with a GUT-scale QCD axion decay constant and a tuned initial misalignment angle to avoid DM overproduction, could be not so predominant in the flux landscape with respect to more natural cases where $f_a\sim \mc{O}(10^{11})$ GeV and $\theta_{\rm in}\sim\mc{O}(\pi)$.

On top of the QCD axion, the isotropic and anisotropic scenarios feature either 1 or 2 ultra-light ALPs. In agreement with previous studies \cite{Mehta:2020kwu, Demirtas:2018akl, Mehta:2021pwf, Halverson:2019cmy}, we argued that the presence of several ultra-light ALPs is a general characteristic of 4D string models where the EFT is under control, as we have shown explicitly in a model with arbitrary $h^{1,1}$ where full moduli stabilisation can be achieved by exploiting higher derivative $\alpha'$ corrections following \cite{Cicoli:2016chb}. Interestingly, we found that the decay constants, the mass spectrum and the contribution to the DM abundance of all these ultra-light ALPs are also logarithmically distributed in the type IIB flux landscape. 

In our recent paper \cite{Broeckel:2020fdz} we found that the the number of flux vacua is also a logarithmic function of the gravitino mass and the supersymmetry breaking scale. Moreover in App. \ref{OtherDistributions} we showed that other quantities relevant for phenomenology, as the moduli masses and the reheating temperature from moduli decay, share the same statistical properties. We are therefore tempted to argue that most, if not all, of the low-energy properties of the string theory landscape seem to obey a logarithmic distribution once moduli stabilisation is properly taken into account. Apart from the particular case of extra dark radiation in Fibre Inflation models, the only exception which we have encountered so far seems to be the supersymmetry breaking scale in KKLT scenarios which might be characterised by a power-law distribution. However its statistical significance is still unclear since this result relies on the assumption that $W_0$ is uniformly distributed \cite{Denef:2004ze} also in the exponentially small regime where however it is very hard to built explicit examples. The only ones which have been constructed so far feature $W_0=0$ and a flat direction at perturbative level \cite{Demirtas:2019sip,Blumenhagen:2020ire, Demirtas:2020ffz}. The flat direction is lifted by non-perturbative physics which generates dynamically an exponentially small $W_0\sim e^{-1/g_s}$. Exploiting the uniform distribution of the string coupling, this relation would again produce a logarithmic distribution of the gravitino mass. 

It is worth stressing that these distributions follow from moduli stabilisation which applies only to corners of the string landscape where the EFT is under control thanks to supersymmetry and weak couplings. In order to judge their genericity one would have therefore to be able to control the EFT beyond the regime of validity of these approximations. Despite the difficulty to achieve this goal, scaling arguments and approximate symmetries inherited from the 10D theory \cite{Burgess:2020qsc} could be used as a powerful guideline to shed light on larger portions of the string landscape. This top-down analysis of the statistical properties of quantities relevant for phenomenology is crucial to provide more theoretical guidance to recent bottom-up approaches to understand naturalness and string theory predictions for several observables \cite{Baer:2017uvn, Baer:2020vad,Baer:2020dri,Baer:2021uxe,Baer:2021vrk,Baer:2021aax}.

We finally comment on the fact that the relative flatness of logarithmic distributions in the string landscape might be seen at first sight as an indication of a difficulty to make sharp predictions from string theory. However a key-feature of string theory is the correlation between different low-energy phenomenological quantities due to the underlying UV framework. It is this interplay which should be used to sharpen the predictions of the string landscape. As an example, we mention the fact that in LVS models an intermediate scale QCD axion decay constant would correlate with TeV-scale soft terms and a volume mode mass around $1$ MeV. Thus, in the absence of a mechanism to avoid cosmological problems associated to the presence of such a light modulus \cite{Kane:2015jia}, a natural QCD axion DM situation with $f_a\sim \mc{O}(10^{11})$ GeV and $\theta_{\rm in}\sim\mc{O}(\pi)$ would not be viable even if the number of vacua with these features is only logarithmically suppressed with respect to the number of vacua with a GUT-scale decay constant. We leave the important study of the UV correlation between different particle physics and cosmological observables with logarithmic distributions for future work.

\section*{Acknowledgements}

We would like to thank Veronica Guidetti and Manki Kim for helpful discussions. The work of KS is supported in part by DOE Grant desc0009956.

\appendix 

\section{Canonical normalisation}
\label{AppCanNorm}

In this appendix we shall perform the canonical normalisation of the axion fields. 

\subsection{A single axion}
\label{fdef}

Let us start with the simple case with a single closed string modulus $T=\tau + {\rm i} \theta$ where it is easy to identify the correct definition of the axion decay constant and periodicity. We start with the following Lagrangian:
\be
\mc{L}=K_{T\bar{T}}\partial_\mu\theta\partial^\mu\theta-\frac14 {\rm Re}(f)\, F^{\mu\nu}_b F_{\mu\nu}^b - \frac14 {\rm Im}(f)\, F^{\mu\nu}_b \tilde{F}_{\mu\nu}^b + \Lambda^4 \cos\left(\frac{2\pi}{\mathfrak{n}}\theta\right),
\ee
where $b$ is a non-Abelian index and the gauge kinetic function is given by $f=T/(2\pi)$. Expressing $\mc{L}$ in terms of the canonically normalised axion $a = \sqrt{2 K_{T\bar{T}}}\,\theta$ and Yang-Mills field strength $G_{\mu\nu}^b = \sqrt{{\rm Re}(f)}\, F^{\mu\nu}_b$, we end up with:
\be
\mc{L}=\frac12\partial_\mu a\partial^\mu a-\frac14 G^{\mu\nu}_b G_{\mu\nu}^b - \frac{\alpha_b}{4} \frac{a}{\sqrt{2 K_{T\bar{T}}}}\, G^{\mu\nu}_b \tilde{G}_{\mu\nu}^b + \Lambda^4 \cos\left(\frac{2\pi}{\mathfrak{n}}\frac{a}{\sqrt{2 K_{T\bar{T}}}}\right),
\ee
where we used the fact that $\tau = \alpha_b^{-1}$. This expression suggests the definition of the axion decay constant $f_a$ as (inserting the appropriate power of $M_p$):
\be
f_a\equiv \left(\frac{\mathfrak{n}}{2\pi}\right)\sqrt{2 K_{T\bar{T}}}\, M_p\,,
\ee
since $\mc{L}$ would simplify to the standard expression:
\be
\mc{L}=\frac12\partial_\mu a\partial^\mu a-\frac14 G^{\mu\nu}_b G_{\mu\nu}^b  - \frac{a}{f_a}\frac{\mathfrak{n}\alpha_b}{8\pi}\, G^{\mu\nu}_b \tilde{G}_{\mu\nu}^b + \Lambda^4 \cos\left(\frac{a}{f_a}\right).
\ee

\subsection{A more general case with 3 axions}

Without loss of generality we shall consider the volume form (\ref{voform}) with $\tau_4=0$. The K\"ahler metric and its inverse take the following form at leading order in a large-$\vo$ expansion:
\be
\mc{K} = 
\begin{pmatrix}
\frac{1}{4\tau_1^2} & \frac{\gamma_3}{4}\frac{\tau_3^{3/2}}{\tau_1^{3/2}\tau_2^2} & -\frac{3\gamma_3}{8}\frac{\sqrt{\tau_3}}{\tau_1^{3/2}\tau_2} \\
\frac{\gamma_3}{4}\frac{\tau_3^{3/2}}{\tau_1^{3/2}\tau_2^2}& \frac{1}{2\tau_2^2} & -\frac{3\gamma_3}{4}\frac{\sqrt{\tau_3}}{\sqrt{\tau_1}\tau_2^2}  \\
-\frac{3\gamma_3}{8}\frac{\sqrt{\tau_3}}{\tau_1^{3/2}\tau_2}  & -\frac{3\gamma_3}{4}\frac{\sqrt{\tau_3}}{\sqrt{\tau_1}\tau_2^2}  & \frac{3\gamma_3}{8}\frac{1}{\sqrt{\tau_3\tau_1}\tau_2}
\end{pmatrix}\,,
\label{fib_Kahler_metr}
\ee
and:
\be
\mc{K}^{-1} = 
\begin{pmatrix}
4\tau_1^2 & 4\gamma_3 \sqrt{\tau_1}\tau_3^{3/2} & 4\tau_1 \tau_3 \\
4\gamma_3 \sqrt{\tau_1}\tau_3^{3/2} & 2\tau_2^2 & 4\tau_2\tau_3 \\
4\tau_1 \tau_3 & 4\tau_2\tau_3 & \frac{8}{3\gamma_3} \sqrt{\tau_3\tau_1} \tau_2
\end{pmatrix}\,.
\label{InvKahler}
\ee
Let us now consider the isotropic and anisotropic limits separately.

\subsubsection*{Isotropic limit}

In the isotropic limit $\theta_1$ and $\theta_2$ are essentially massless while $\theta_3$ develops a potential via QCD instantons of the form $V(\theta_3) = -\Lambda_\QCD^4 \cos(2\pi\theta_3)$. Hence the only non-zero entry of the axionic Hessian is $V_{33}= (2\pi)^2 \Lambda_\QCD^4$. Multiplying the inverse K\"ahler metric (\ref{InvKahler}) by the axionic Hessian we find the mass-squared matrix $\mc{M}^2=\frac12\mc{K}^{-1}V_{ij}$ which becomes:
\be
\mc{M}^2 = 
\begin{pmatrix}
0 & 0 & 2  \tau_1\tau_3 \\
0  & 0 & 2  \tau_2\tau_3 \\
0 & 0 & \frac{4}{3\gamma_3}\sqrt{\tau_3\tau_1}\tau_2
\end{pmatrix} (2\pi)^2 \Lambda_\QCD^4\,.
\label{fib_massmatrix}
\ee
The eigenvalues of $\mc{M}^2$ are (reinstating appropriate powers of $M_p$):
\be
m_1^2=0\,,\qquad m_2^2=0\,,\qquad m_3^2 = \frac{4}{3\gamma_3}\sqrt{\tau_3\tau_1}\tau_2 (2\pi)^2 \frac{\Lambda_\QCD^4}{M_p^2}\,,
\label{AxMass}
\ee
and the corresponding eigenvectors read:
\be
\vec{v}_1 = 
\begin{pmatrix}
1 \\ 0 \\ 0
\end{pmatrix}\frac{\mathfrak{n}_1}{2\pi}\frac{M_p}{f_1}\,,
\qquad
\vec{v}_2 = 
\begin{pmatrix}
0 \\ 1 \\ 0
\end{pmatrix}\frac{\mathfrak{n}_2}{2\pi}\frac{M_p}{f_2}\,,
\qquad
\vec{v}_3 = 
\begin{pmatrix}
\frac{3\gamma_3}{2}\frac{\sqrt{\tau_3\tau_1}}{\tau_2} \\ \frac{3\gamma_3}{2}\sqrt{\frac{\tau_3}{\tau_1}} \\ 1
\end{pmatrix}\frac{M_p}{2\pi f_3}\,,
\ee
where $f_1$, $f_2$ and $f_3$ are the axion decay constants which can be obtained by requiring $\vec{v}_i^\T\mc{K}\vec{v}_j=\frac12 \delta_{ij}$. We find (at leading order in a large-$\vo$ approximation):
\be
f_1=\frac{\mathfrak{n}_1}{2\pi}\frac{M_p}{\sqrt{2}\tau_1}\,,\qquad f_2=\frac{\mathfrak{n}_2}{2\pi}\frac{M_p}{\tau_2}\,,\qquad
f_3 = \frac{\sqrt{3\alpha \gamma_3}}{4\pi}\frac{M_p}{\tau_3^{1/4}\sqrt{\vo}}\,.
\ee
Therefore the QCD axion mass in (\ref{AxMass}) can correctly be written also as $m_3 = \Lambda_\QCD^2/f_3$. Moreover the original axions $\theta_i$'s can be expressed in terms of the canonically normalised axions $a_i$'s as:
\bea
\theta_1 &=& \frac{\mathfrak{n}_1}{2\pi}\frac{a_1}{f_1} + \frac{3\gamma_3}{4\pi}\frac{\sqrt{\tau_3\tau_1}}{\tau_2} \frac{a_3}{f_3} \,, \nn \\
\theta_2 &=& \frac{\mathfrak{n}_2}{2\pi}\frac{a_2}{f_2} +
\frac{3\gamma_3}{4\pi}\sqrt{\frac{\tau_3}{\tau_1}} \frac{a_3}{f_3} \,, \\
\theta_3 &=& \frac{1}{2\pi}\frac{a_3}{f_3}\,. \nn
\eea

\subsubsection*{Anisotropic limit}

In the anisotropic limit $\theta_3$ develops a potential via non-perturbative corrections to $W$, $\theta_2$ is essentially massless while $\theta_1$ becomes massive via QCD instantons. Hence the axionic Hessian at the minimum takes the form:
\be
V_{ij} = 
\begin{pmatrix}
(2\pi)^2 \Lambda_\QCD^4 & 0 & 0 \\
0  & 0 & 0 \\
0 & 0 & 3\alpha \mathfrak{a}_3^2 \tau_3^{3/2}\frac{W_0^2}{\vo^3}
\end{pmatrix}
\label{Hessian}.
\ee
Thus the mass-squared matrix $\mc{M}^2=\frac12\mc{K}^{-1}V_{ij}$ now becomes:
\be
\mc{M}^2 = 
\begin{pmatrix}
8\pi^2\tau_1^2 \Lambda_\QCD^4  & 0 & 6\alpha \mathfrak{a}_3^2 \tau_1\tau_3^{5/2}\frac{W_0^2}{\vo^3} \\
8\gamma_3\pi^2 \sqrt{\tau_1}\tau_3^{3/2} \Lambda_\QCD^4  & 0 & 6 \mathfrak{a}_3^2\tau_3^{5/2}\tau_1^{-1/2}\frac{W_0^2}{\vo^2} \\
8\pi^2\tau_1\tau_3 \Lambda_\QCD^4  & 0 & 4 \mathfrak{a}_3^2 \gamma_3^{-1}\tau_3^2\frac{W_0^2}{\vo^2}
\end{pmatrix}\,.
\label{fib_massmatrixAnis}
\ee
The leading order expressions of the eigenvalues of $\mc{M}^2$ are (inserting suitable powers of $M_p$):
\be
m_1^2=8\pi^2 \tau_1^2 \,\frac{\Lambda_\QCD^4}{M_p^2} \,,\qquad m_2^2=0\,,\qquad m_3^2 = \frac{4 \mathfrak{a}_3^2 \tau_3^2}{\gamma_3}\left(\frac{W_0}{\vo}\right)^2 M_p^2 \,,
\label{AxMassAniso}
\ee
and the corresponding eigenvectors read:
\be
\vec{v}_1 = 
\begin{pmatrix}
1 \\ -\frac{\gamma_3}{2} \left(\frac{\tau_3}{\tau_1}\right)^{3/2} \\
- \frac{\gamma_3\mathfrak{n}_3^2}{2}\frac{\tau_1}{\tau_3}\frac{\vo^2}{W_0^2}\Lambda_\QCD^4
\end{pmatrix}\frac{\mathfrak{n}_1}{2\pi}\frac{M_p}{f_1}\,,
\qquad
\vec{v}_2 = 
\begin{pmatrix}
0 \\ 1 \\ 0
\end{pmatrix}\frac{\mathfrak{n}_2}{2\pi}\frac{M_p}{f_2}\,,
\qquad
\vec{v}_3 = 
\begin{pmatrix}
\frac{3\alpha\gamma_3}{2}\frac{\sqrt{\tau_3}\tau_1}{\vo} \\ \frac{3\gamma_3}{2}\sqrt{\frac{\tau_3}{\tau_1}} \\ 1
\end{pmatrix}\frac{\mathfrak{n}_3}{2\pi}\frac{M_p}{f_3}\,, 
\ee
where $f_1$, $f_2$ and $f_3$ are the axion decay constants which can be obtained by requiring $\vec{v}_i^\T\mc{K}\vec{v}_j=\frac12 \delta_{ij}$. We obtain (at leading order in a large-$\vo$ approximation):
\be
f_1=\frac{1}{2\sqrt{2}\pi}\frac{M_p}{\tau_1}\,,\qquad f_2=\frac{\mathfrak{n}_2}{2\pi}\frac{M_p}{\tau_2}\,,\qquad
f_3 = \frac{\mathfrak{n}_3\sqrt{3\alpha \gamma_3}}{4\pi}\frac{M_p}{\tau_3^{1/4}\sqrt{\vo}}\,.
\ee
Therefore the QCD axion mass in (\ref{AxMassAniso}) can correctly be written also as $m_1 = \Lambda_\QCD^2/f_1$. Moreover the original axions $\theta_i$'s can be expressed in terms of the canonically normalised axions $a_i$'s as:
\bea
\theta_1 &=& \frac{1}{2\pi}\frac{a_1}{f_1} + \frac{3\alpha\mathfrak{n}_3\gamma_3}{4\pi}\frac{\sqrt{\tau_3}\tau_1}{\vo} \frac{a_3}{f_3}\,,  \nn \\
\theta_2 &=& -\frac{\gamma_3}{4\pi} \left(\frac{\tau_3}{\tau_1}\right)^{3/2}\frac{a_1}{f_1}+\frac{\mathfrak{n}_2}{2\pi}\frac{a_2}{f_2} +
\frac{3\mathfrak{n}_3\gamma_3}{4\pi}\sqrt{\frac{\tau_3}{\tau_1}} \frac{a_3}{f_3}\,,  \\
\theta_3 &=& - \frac{1}{2\pi}\frac{\tau_3}{\tau_1}\left(\frac{m_1}{m_3}\right)^2\frac{a_1}{f_1}+\frac{\mathfrak{n}_3}{2\pi}\frac{a_3}{f_3}\,. \nn
\eea

\section{Benchmark points for ALP dark matter}
\label{abun}

In this appendix we present some benchmark points for ALP DM generated by the misalignment mechanism. In this case the DM relic abundance is given by (\ref{nonQCDCDM}). We focus on the bulk axion $\theta_2$ which behaves as an ultra-light ALP for both the isotropic and the anisotropic case. Its mass and decay constant can be written in terms of the underlying parameters as:
\be
m_{\theta_2}= \sqrt{\frac12 \mc{K}^{-1}_{22} V_{\theta_2\theta_2}} = \frac{4\pi}{\alpha\mathfrak{n}_2^{3/2}} 
\,\sqrt{g_s A_2 W_0}\,\sqrt{\frac{\tau_2}{\tau_1}}\, e^{-\frac{\pi}{\mathfrak{n}_2}\tau_2}\,M_p\,, \qquad
f_{\theta_2}=\frac{\mathfrak{n}_2}{2\pi}\frac{M_p}{\tau_2}\,.
\label{m2f2}
\ee
In the isotropic case $\tau_1 = \tau_2$, while in the anisotropic limit $\tau_1 = g_s^2\,\tau_2 = \alpha_\SM^{-1}$. After writing $\tau_1$ in terms of $\tau_2$, $\tau_2$ can in turn be expressed as a function of the microscopic parameters using (\ref{swiss_minimum}) with $\vo\simeq \alpha\sqrt{\tau_1}\tau_2$. The expression (\ref{nonQCDCDM}) for the ALP DM abundance becomes then just a function of 9 UV parameters: $g_s$, $W_0$, $\mathfrak{n}_2$, $\mathfrak{n}_4$, $A_2$, $A_4$, $\alpha$, $\xi$ and $\theta_{2,{\rm in}}$. In what follows we shall restrict our numerical search for benchmark examples to a 4D subregion of this parameter space by focusing on natural values $A_2=A_4=1$ and $\theta_{2,{\rm in}}=\pi$. Moreover we shall set the topological quantities $\alpha=1/6$ and $\xi = 0.46$ as in the explicit toric constructions of \cite{Cicoli:2016xae}. In Tables \ref{Table1} and \ref{Table2} we present some benchmark points which reproduce the observed DM abundance for different values of $W_0$, $g_s$, $\mathfrak{n}_2$ and $\mathfrak{n}_4$, for the isotropic and anisotropic cases respectively. 

\begin{table}
\centering
\begin{tabular}{ | p{0.5cm} | p{0.5cm} | p{1.0cm} | p{1.9cm} || p{1.2cm} |  p{2.2cm} | p{2.2cm} | }
\hline
$\mathfrak{n}_2$ & $\mathfrak{n}_4$ & $\quad g_s$ & $\quad\,\, W_0$ & $\quad\tau_2$  & $\,\,\, m_{\theta_2}$ (eV) & $\,\, f_{\theta_2}$ (GeV) \\
\hline
  $1$ & $1$ & $0.250$  & $2.05\cdot 10^{-11}$ & $\,32.39$ & $\,\, 2.62\cdot10^{-21}$ & $\,\, 1.18\cdot10^{16}$ \\
    $1$ & $1$ & $0.100$ & $4.46\cdot 10^{-32}$ & $\,24.97$ & $\,\, 1.02\cdot10^{-21}$ & $\,\, 1.53\cdot10^{16}$ \\
     $1$ & $1$ & $0.075$ & $1.58\cdot 10^{-43}$ & $\,20.97$ & $\,\, 4.79\cdot10^{-22}$ & $\,\, 1.82\cdot10^{16}$ \\
    $1$ & $10$ & $0.250$ & $3.66$ & $\,36.36$  & $\,\, 4.29\cdot10^{-21}$ & $\,\, 1.05\cdot10^{16}$ \\
  $1$ & $10$ & $0.100$ & $2.08\cdot 10^{-2}$ & $\,35.42$  & $\,\, 3.87\cdot10^{-21}$ & $\,\, 1.08\cdot10^{16}$ \\
   $1$ & $10$ & $0.075$ & $1.32\cdot 10^{-3}$ & $\,34.96$  & $\,\, 3.61\cdot10^{-21}$ & $\,\, 1.09\cdot10^{16}$ \\
  \hline
    $10$ & $1$ & $0.250$ & $6.32\cdot 10^{-10}$ & $318.47$ & $\,\, 2.54\cdot10^{-21}$ & $\,\, 1.20\cdot10^{16}$ \\
    $10$ & $1$ & $0.100$ & $1.37\cdot 10^{-30}$ & $244.35$ & $\,\, 9.70\cdot10^{-22}$ & $\,\, 1.56\cdot10^{16}$ \\
  $10$ & $1$ & $0.075$ & $4.81\cdot 10^{-42}$ & $204.47$ & $\,\, 4.35\cdot10^{-22}$ & $\,\, 1.87\cdot10^{16}$ \\  
  $10$ & $10$ & $0.250$ & $113.2$ & $358.27$  & $\,\, 4.00\cdot10^{-21}$ & $\,\, 1.07\cdot10^{16}$ \\
  $10$ & $10$ & $0.100$ & $0.643$ & $348.90$  & $\,\, 3.62\cdot10^{-21}$ & $\,\, 1.09\cdot10^{16}$ \\
     $10$ & $10$ & $0.075$ & $4.07\cdot 10^{-2}$ & $344.27$  & $\,\, 3.37\cdot10^{-21}$ & $\,\, 1.11\cdot10^{16}$ \\
 \hline
\end{tabular}
\caption{Benchmark points which match the observed ALP DM abundance for the isotropic case setting $A_2=A_4=1$, $\theta_{{\rm in},2}=\pi$, $\alpha=1/6$ and $\xi=0.46$.}
\label{Table1}
\end{table}

\begin{table}
\centering
\begin{tabular}{ | p{0.5cm} | p{0.5cm} | p{1.0cm} | p{1.9cm} || p{1.2cm} |  p{2.2cm} | p{2.2cm} | }
\hline
$\mathfrak{n}_2$ & $\mathfrak{n}_4$ & $\quad g_s$ & $\quad\,\, W_0$ & $\quad\tau_2$  & $\,\,\, m_{\theta_2}$ (eV) & $\,\, f_{\theta_2}$ (GeV) \\
\hline
  $1$ & $1$ & $0.833$ & $0.108$ & $\,36.03$ & $\,\, 4.53\cdot10^{-21}$ & $\,\, 1.06\cdot10^{16}$ \\
  $1$ & $10$ & $0.822$ & $48.93$ & $\,37.00$ & $\,\, 4.61\cdot10^{-21}$ & $\,\, 1.03\cdot10^{16}$ \\
\hline
  $6$ & $1$ & $0.352$ & $1.11\cdot 10^{-6}$ & $201.66$ & $\,\, 3.06\cdot10^{-21}$ & $\,\, 1.14\cdot10^{16}$  \\
  $6$ & $10$ & $0.339$ & $47.88$ & $217.80$ & $\,\, 4.36\cdot10^{-21}$ & $\,\, 1.05\cdot10^{16}$ \\
\hline
$10$ & $1$ & $0.277$ & $4.05\cdot 10^{-9}$ & $325.35$ & $\,\, 2.82\cdot10^{-21}$ & $\,\, 1.17\cdot10^{16}$  \\
 $10$  & $10$ & $0.263$ & $36.15$ & $360.61$ & $\,\, 4.22\cdot10^{-21}$ & $\,\, 1.06\cdot10^{16}$ \\
  \hline
  $30$ & $1$ & $0.168$ & $1.06\cdot 10^{-16}$ & $883.67$ & $\,\, 1.79\cdot10^{-21}$ & $\,\, 1.30\cdot10^{16}$ \\
 $30$ & $10$ & $0.153$ & $9.75$ & $1063.23$ & $\,\, 3.88\cdot10^{-21}$ & $\,\, 1.08\cdot10^{16}$ \\
  \hline
\end{tabular}
\caption{Benchmark points which match the observed ALP DM abundance for the anisotropic case setting $A_2=A_4=1$, $\theta_{{\rm in},2}=\pi$, $\alpha=1/6$ and $\xi=0.46$. All benchmark points satisfy the phenomenological constraint $\tau_1=\alpha_\SM^{-1}=25$.}
\label{Table2}
\end{table}

Notice that in both cases the typical values of the mass and the decay constant are respectively $m_{\theta_2} \simeq 5\cdot 10^{-21}$ eV and $f_{\theta_2} \simeq 10^{16}$ GeV. In the isotropic case we have chosen $\mathfrak{n}_2$, $\mathfrak{n}_4$ and $g_s$ freely (focusing on values of $g_s$ which keep perturbation theory under control) and we have derived the value of $W_0$ which matches the observed DM abundance. Notice that natural $\mc{O}(1$-$10)$ values of $W_0$ require $\mathfrak{n}_4\gtrsim 10$ since from (\ref{swiss_minimum}) $\tau_2 \sim W_0^{2/3}\,e^{k/\mathfrak{n}_4}$ for an appropriate $k$, and so $\mathfrak{n}_4\sim \mc{O}(1)$ would give a value of $\tau_2$ which is too large to match $\Omega_\ALP h^2\simeq 0.112$ due to the exponential suppression 
$m_{\theta_2} \sim e^{-\frac{\pi}{\mathfrak{n}_2}\tau_2}\,M_p$ 
in (\ref{m2f2}). This relation explains also why larger values of $\tau_2$ correspond to larger values of $\mathfrak{n}_2$. Let us finally stress that in the anisotropic case we have chosen freely only $\mathfrak{n}_2$ and $\mathfrak{n}_4$ since $g_s$ is fixed by the phenomenological constraint $\tau_1=g_s^2\, \tau_2 = \alpha_\SM^{-1}=25$. As can be seen from Table \ref{Table2}, this condition tends to push the string coupling close to $1$ unless $\mathfrak{n}_2\gtrsim 10$ since $g_s^2 \tau_2 = 25$ can be satisfied for $g_s\simeq 0.1$ only for large values of $\tau_2$ which, as we have already pointed out, need large values of $\mathfrak{n}_2$.

\section{Other distributions relevant for phenomenology}
\label{OtherDistributions}

In this appendix we shall show that other phenomenologically interesting quantities feature also a logarithmic distribution in the type IIB flux landscape. 

\subsection{Moduli masses}

Let us investigate the distribution of moduli masses in the flux landscape. For all cases, the isotropic and anisotropic cases with $h^{1,1}=4$ and the model with arbitrarily large $h^{1,1}$, the mass of each K\"ahler modulus scales with the CY volume as:
\be
m_{\tau_i} \simeq \frac{W_0}{\vo^{p_i}}\,M_p\qquad\text{with}\quad p_i>0\quad \forall i=1,...,h^{1,1}.
\ee
Following the same logic as in Sec. \ref{sec:Pheno}, we find again a logarithmic distribution for each $m_{\tau_i}$ since these masses are controlled by the exponentially large volume $\vo$:
\be
N(m_{\tau_i}) \sim  \ln\left( \frac{m_{\tau_i}}{M_p} \right),\qquad \forall i=1,...,h^{1,1}.
\label{Nmi}
\ee
For the anisotropic case one has just to make sure that the bound $g_s\gtrsim 0.01$ (coming from the ability to tune $W_0$ to satisfy (\ref{W0gsRel})) does not set a lower bound on $m_{\tau_i}$ for the regime of validity of the distribution (\ref{Nmi}). However this bound is negligible since combining (\ref{swiss_minimum}) with (\ref{W0gsRel}) one would find $m_{3/2}\gtrsim 10^{-45}\,M_p\simeq 10^{-16}$ eV for $g_s\gtrsim 0.01$.

\subsection{Reheating temperature}

Using the moduli masses we can study the distribution of the reheating temperature coming from moduli decay \cite{Cicoli:2018cgu, Cicoli:2010ha}. The reheating temperature due to the perturbative decay of the $i$-th K\"ahler modulus is given by \cite{Cicoli:2018cgu}:
\be
T_{{\rm rh},i} = \left( \frac{40 c_{\rm vis} c_{\text{tot}}}{\pi^2g_*(T_{\rm rh})} \right)^{1/4}\sqrt{\Gamma_{\tau_i} M_p}\,,
\label{reheating_temp}
\ee
where $c_{\rm vis}$ and $c_{\rm vis}$ control the strength the interaction of the modulus $\tau_i$ with the visible and the hidden sector respectively, $c_{\rm tot}=c_{\rm vis}+c_{\rm hid}$, and the decay rate $\Gamma_{\tau_i}$ looks like:
\be
\Gamma_{\tau_i} = \frac{1}{48\pi}\frac{m^3_{\tau_i}}{M_p^2}\,.
\ee
Thus using (\ref{Nmi}) we obtain again a logarithmic distribution for all cases:
\be
T_{{\rm rh},i}\sim m_{\tau_i} \sqrt{\frac{m_{\tau_i}}{M_p}}\quad\Rightarrow\quad 
\frac{d T_{{\rm rh},i}}{T_{{\rm rh},i}}\sim \frac{d m_{\tau_i}}{m_{\tau_i}}\quad\Rightarrow\quad
N (T_{\rm rh}) \sim \ln\left( \frac{T_{\rm rh}}{M_p} \right).
\ee

\bibliographystyle{utphys}

\bibliography{mybib}

\end{document}